\documentclass[fleqn,usenatbib]{mnras}

\usepackage{newtxtext,newtxmath}

\usepackage[T1]{fontenc}

\DeclareRobustCommand{\VAN}[3]{#2}
\let\VANthebibliography\thebibliography
\def\thebibliography{\DeclareRobustCommand{\VAN}[3]{##3}\VANthebibliography}

\usepackage{graphicx}	
\usepackage{amsmath}	
\usepackage{booktabs,longtable}
\usepackage{xtab,afterpage}
\usepackage{adjustbox}
\extrafloats{200}
\usepackage{caption}
\usepackage{soul}

\title[MPTA First Data Release]{The MeerKAT Pulsar Timing Array: First Data Release}

\author[M.~T.~Miles et al.]{
M.~T.~Miles,$^{1,2}$\thanks{E-mail: \href{mailto:matthewmiles@swin.edu.au}{matthewmiles@swin.edu.au}}
R.~M.~Shannon,$^{1,2}$
M.~Bailes,$^{1,2}$
D.~J.~Reardon,$^{1,2}$
M.~J.~Keith,$^{3}$
A.~D.~Cameron,$^{1,2}$ \newauthor
A.~Parthasarathy,$^{4}$
M.~Shamohammadi,$^{1,2}$ 
R.~Spiewak,$^{3,1,2}$
W.~van~Straten,$^{5}$
S.~Buchner,$^{6}$
F.~Camilo,$^{6}$ \newauthor
M.~Geyer,$^{6,7}$
A.~Karastergiou,$^{8}$
M.~Kramer,$^{4}$
M.~Serylak,$^{9,10}$ 
G.~Theureau,$^{11,12}$ and
V.~Venkatraman~Krishnan$^{4}$
\\
$^{1}$Centre for Astrophysics and Supercomputing, Swinburne University of Technology, PO Box 218, Hawthorn, VIC 3122, Australia\\
$^{2}$ARC Centre of Excellence for Gravitational Wave Discovery (OzGrav), Mail H29, Swinburne University of Technology, PO Box \\ 218, Hawthorn, VIC 3122, Australia \\
$^{3}$Jodrell Bank Centre for Astrophysics, Department of Physics and Astronomy, University of Manchester, Manchester M13 9PL, UK\\
$^{4}$Max-Planck-Institut f\"{u}r Radioastronomie, Auf dem H\"{u}gel 69, D-53121 Bonn, Germany \\
$^{5}$Institute for Radio Astronomy \& Space Research, Auckland University of Technology, Private Bag 92006, Auckland 1142, New Zealand \\
$^{6}$South African Radio Astronomy Observatory, 2 Fir Street, Black River Park, Observatory 7925, South Africa \\
$^{7}$Department of Astronomy, University of Cape Town, Rondebosch, Cape Town, 7700, South Africa \\
$^{8}$Department of Astrophysics, University of Oxford, Denys Wilkinson Building, Keble Road, Oxford OX1 3RH, UK \\
$^{9}$SKA Observatory, Jodrell Bank, Lower Withington, Macclesfield, SK11 9FT, United Kingdom \\
$^{10}$Department of Physics and Astronomy, University of the Western Cape, Bellville, Cape Town, 7535, South Africa \\
$^{11}$Laboratoire de Physique et Chimie de l'Environnement et de l'Espace, LPC2E UMR7328, Université d'Orléans, CNRS, F-45071 Orléans, France \\
$^{12}$Laboratoire Univers et Théories, Observatoire de Paris, Université PSL, Université de Paris Cité, CNRS, F-92190 Meudon, France \\
}

\date{Accepted XXX. Received YYY; in original form ZZZ}

\pubyear{2022}

\begin{document}
\label{firstpage}
\pagerange{\pageref{firstpage}--\pageref{lastpage}}
\maketitle

\begin{abstract}
We present the first 2.5 years of data from the MeerKAT Pulsar Timing Array (MPTA), part of MeerTime, a MeerKAT Large Survey Project. The MPTA aims to precisely measure pulse arrival times from an ensemble of 88 pulsars visible from the Southern Hemisphere, with the goal of contributing to the search, detection and study of nanohertz-frequency gravitational waves as part of the International Pulsar Timing Array. This project makes use of the MeerKAT telescope, and operates with a typical observing cadence of two weeks using the L-band receiver that records data from 856-1712 MHz. We provide a comprehensive description of the observing system, software, and pipelines used and developed for the MeerTime project. The data products made available as part of this data release are from the 78 pulsars that had at least $30$ observations between the start of the MeerTime programme in February 2019 and October 2021. These include both sub-banded and band-averaged arrival times, as well as the initial timing ephemerides, noise models, and the frequency-dependent standard templates (portraits) used to derive pulse arrival times. After accounting for detected noise processes in the data, the frequency-averaged residuals of $67$ of the pulsars achieved a root-mean-square residual precision of $< 1 \mu \rm{s}$. We also present a novel recovery of the clock correction waveform solely from pulsar timing residuals, and an exploration into preliminary findings of interest to the international pulsar timing community. The arrival times, standards and full Stokes parameter calibrated pulsar timing archives are publicly available.
\end{abstract}




\section{Introduction}
Millisecond pulsars (MSPs) are a subclass of pulsars that are thought to have experienced a unique evolutionary path, involving the accretion of material from a binary companion. The subsequent transfer of angular momentum from this process results in a rotational frequency up to a factor of a thousand greater than would be expected of an ordinary pulsar, 
a reduction in their magnetic field strength down to $\sim 10^8$ G, and makes them far more rotationally stable. Observations of individual members of this subclass have proven incredibly fruitful leading to extensive tests of general relativity \citep{2012MNRAS.423.3328F, 2018Natur.559...73A, 2020A&A...638A..24V}, the discovery of planetary systems hosted by distant stars \citep{Wolszczan_Frail_1992, Wolszczan_planets_1994}, and allowing us to place ever-tightening constraints on nuclear equations of state \citep{Demorest_Nature_2010, Antoniadis1233232, 2021ApJ...915L..12F, 2021arXiv210506979M, 2021arXiv210506980R}. The  precision and stability of these pulsars enables the realisation of sensitive pulsar timing arrays (PTAs), ensembles of pulsars that are consistently observed, and their pulse times of arrival (ToAs) catalogued and measured for up to decades \citep{1978SvA....22...36S, 1979ApJ...234.1100D, 1990ApJ...361..300F}. The primary goal of PTA observations is the detection of nanohertz-frequency gravitational radiation \citep{1990ApJ...361..300F}. The most likely first source of nanohertz-frequency gravitational waves (GWs) is a stochastic gravitational-wave background (SGWB) that is the superposition of many sources \citep{Hellings_Downs_1983}. Several recent searches show potential evidence for a common spectral noise process that could indicate the presence of such a background \cite[][]{NanoGravGWB, 2021arXiv210712112G, 2021MNRAS.508.4970C, 2022MNRAS.510.4873A}, although it remains to be seen if this common red noise process would follow the expected Hellings-Downs (HD) angular correlation. The most likely progenitors of the GW signals in this frequency-band are thought to be inspiralling supermassive black-hole binaries (SMBHBs) \citep{1995ApJ...446..543R, 2003ApJ...583..616J, 2003ApJ...590..691W, 2004ApJ...611..623S, 2011MNRAS.411.1467K, 2012A&A...545A.127R, 2017PhRvL.118r1102T}. However, there may exist additional sources from which a SGWB may arise, such as from cosmic strings \citep{1976JPhA....9.1387K, PhysRevD.81.104028, 2012PhRvD..85l2003S, 2015MNRAS.453.2576L, 2018ApJ...859...47A}, or cosmological phase transitions \citep{1980PhLB...91...99S, 2005PhyU...48.1235G}.
 
The MeerKAT radio telescope\footnote{Operated by the South African Radio Astronomy Observatory (SARAO).} is a $64$-antenna interferometer located in the Northern Karoo region of South Africa, and is the precursor to the upcoming Square Kilometer Array Mid telescope \citep[SKA;][]{2009IEEEP..97.1482D} demonstrating SKA technology on a SKA site. Millisecond pulsar timing with MeerKAT (The MeerKAT Pulsar Timing Array\footnote{Sometimes referred to as the MeerTime Pulsar Timing Array.}, MPTA), is undertaken as part of the MeerTime Large Survey Project (LSP) \citep{2016mks..confE..11B}. The potential of the MeerKAT PTA project has already been demonstrated through dedicated studies of jitter noise in the MSP population \citep{2021MNRAS.502..407P}, a comprehensive census of the MSPs visible to MeerKAT and their emission properties \citep{spiewak2022}, and the discovery of mode-changing in one of the most precisely timed and often observed millisecond pulsars \citep{2022MNRAS.510.5908M}.

The MPTA is poised to make a significant contribution to the International Pulsar Timing Array \cite[IPTA;][]{2010CQGra..27h4013H}.
The IPTA is currently a collaboration between four individual PTAs (chronologically, the European PTA \citep[EPTA;][]{2008AIPC..983..633J}, the Parkes PTA \citep[PPTA;][]{2008AIPC..983..584M}, the North American Nanohertz Observatory for Gravitational Waves \citep[NANOGrav;][]{2009arXiv0909.1058J}, and the Indian PTA \citep[InPTA;][]{2022arXiv220609289T}) to search for low-frequency gravitational waves. Of the major PTAs that are members of this effort, the PPTA has so far been the only one capable of observing the entire Southern sky. The MPTA can therefore make a key contribution to the observation of Southern MSPs.
MeerKAT has shown itself to be many times more sensitive than the Parkes ({\em Murriyang}) radio telescope, possessing an array gain that is $\sim 4$ times greater, combined with a lower system temperature further bolstering its sensitivity. 
In addition to this high sensitivity MeerKAT's ability to quickly slew between sources and the access it has to a large swathe of pulsars (Dec $<+44^{\circ}$), makes the MPTA naturally suited to contribute to the IPTA through high precision timing of a large array of pulsars. Forecasts show that the MeerKAT PTA will be contributing roughly half of the IPTA sensitivity by 2025 with current target lists and cadences \cite[][]{spiewak2022}.

In this paper, we present the initial 2.5-year MPTA data release. These data include both sub-banded and frequency averaged ToAs, the full data archives used to construct this data release, the ephemerides that have been used to perform timing, and band-averaged and sub-banded timing residuals. Also included are both the timing standards (frequency averaged, one dimensional) and frequency resolved standards (portraits), used to calculate each set of ToAs as well as the preliminary noise models constructed to facilitate this work. This is the first such data release from the MPTA suitable for obtaining arrival times. In Section \ref{Section: pulsars, obs, data reduction} we describe the pulsars, observations taken, and the process of data reduction. In Section \ref{Section: data set / data products} we present the final data products. In Section \ref{Section: Discussion} we discuss the performance of the data and explore various features and investigations that are contained in this data release. Finally, in Section \ref{Section: Summary and Conclusions} we conclude and summarise our findings.

\section{Pulsar Selection, Observations, and Data Reduction}
\label{Section: pulsars, obs, data reduction}

This data release comprises observations taken exclusively at the MeerKAT radio telescope, an interferometer consisting of $64$, $13.9$-m dishes, located in the Great Karoo region of South Africa. While MeerKAT currently possesses L-band ($856 - 1712$ MHz) and Ultra-High Frequency UHF ($544 - 1088$ MHz) receivers, all data collected for this project used the L-band system, following its successful commissioning. We have not collected any data with the UHF receiver which was commissioned later. In addition, we expect observations in L-band to be less impacted by noise introduced by interstellar medium (ISM) propagation effects, even though UHF observations potentially provide more accurate measurements of dispersion measure \citep{2010arXiv1010.3785C}. MeerKAT possesses a gain of $2.8$ K\,$\mathrm{Jy}^{-1}$, and a system temperature of $\sim 18$K in this band, which results in a system equivalent flux density of $\sim 7$\,Jy. The data were recorded with the SKA prototype pulsar processors, the pulsar timing user-supplied equipment, collectively known as PTUSE \citep{2020PASA...37...28B}.

\subsection{Selection Criteria and Observing Strategy}
\label{Subsection: selection crit and obs strat}

\begin{figure*}
    
    \includegraphics[width=\textwidth]{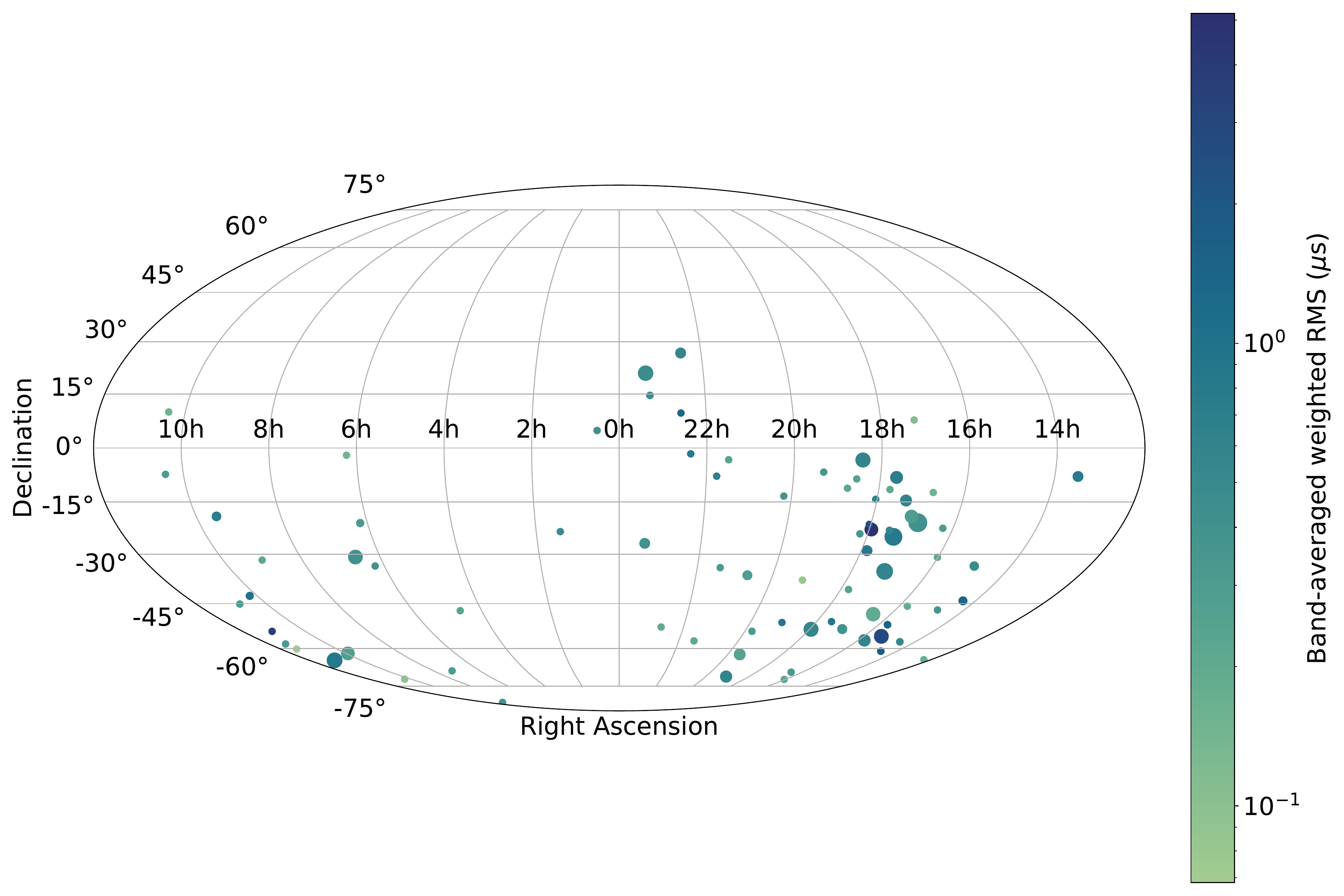}
    \caption[]{MPTA pulsars included in this data release. The color of each pulsar refers to the weighted root-mean-square error of its band-averaged residuals. As most pulsar errors are sub-microsecond, the color is displayed in a log-scale to more easily contrast them. The size of the pulsars refers to their relative median observation times (e.g. larger circles indicate pulsars with typically longer observations).}
    \label{fig: skymap}
\end{figure*}

The data presented here are taken from shortly after the MeerTime project began, spanning from February 2019 through to October 2021 (MJD $58526 - 59510$). This provides approximately $2.5$ years of data for the majority of the pulsars included in the MPTA. The full MPTA includes $88$ pulsars, which have been selected from a census of $189$ MSPs visible to MeerKAT \citep{spiewak2022}. The pulsars were initially chosen to obtain a band averaged timing precision of $<1 \mu\mathrm{s}$ within an integration time of $1000$ seconds. Since the initial selection of the MPTA pulsars, the list has been regularly reviewed, and pulsars have been added and removed from regular timing to maximise the efficiency of the PTA. Similarly, some pulsars have not been included in the MPTA, despite possessing high timing precision, as they are only observable at  declinations that are not ideal for the MeerKAT telescope, so better observed with Northern hemisphere telescopes. The  strategy for the pulsars that are currently included in the MPTA is to regularly observe with a minimum cadence of two weeks per pulsar. There is a minimum integration time of $256$ seconds per observation, if the pulsar is able to achieve sub-microsecond timing precision in that time. Where this cannot be achieved, the pulsar is observed for an integration time required for sub-microsecond precision, if this can be achieved within $2048$ s. While the current MPTA includes $88$ pulsars, this data release includes only the $78$ pulsars that had at least $30$ observations over this observing span. The positions of the pulsars are presented in Figure \ref{fig: skymap}. All of the MSPs experience interstellar scintillation that amplifies and deamplifies their flux densities. The effects of this are more dramatic at low dispersion measures and means that sometimes the pulsars do not reach their target timing residual in their allocated integration times.

While most of the observations follow this two week cadence, there are some exceptions. Notably, $38$ of the binary pulsars in the MPTA have been observed by the MeerTime relativistic binary programme \citep{2021MNRAS.504.2094K} which includes longer observations and single orbit observing campaigns to precisely measure Shapiro delays, enabling companion and pulsar masses to be derived along by determining the orbital inclination angle.

A summary of ten of the pulsars that are included in this data release is provided in Table \ref{table: MPTA_summary}, with the full list included in Table \ref{table: MPTA_summary_full_table}. Included in this table are the pulse period ($P$), the dispersion measure (DM), the orbital period where applicable ($\textrm{P}_{\textrm{b}}$), and the median, mean, and standard deviation of the sub-banded ToA uncertainties. Also included is the frequency-averaged effective pulse width ($\mathrm{W_{eff}}$) \citep{2010arXiv1010.3785C, 2021MNRAS.502..407P}, which we discuss in Section \ref{Subsection: Data set performance}. This is presented as it is expected to provide a better prediction of timing precision than more conventional width measurements.

\begin{table*}

\caption{MPTA partial parameters and timing statistics for the first 10 pulsars in order of Right Ascension. The full description of these parameters can be found at Table \ref{table: MPTA_summary_full_table}.}
\label{table: MPTA_summary}
\begin{tabular*}{\textwidth}[c]{@{\extracolsep{\fill}} lcccccccc}
\toprule
Pulsar &
  \multicolumn{1}{c}{P} &
  \multicolumn{1}{c}{DM} &
  \multicolumn{1}{c}{$\textrm{P}_{\textrm{b}}$} &
  \multicolumn{1}{c}{$\textrm{W}_{\textrm{eff}}$} &
  \multicolumn{3}{c}{Uncertainty ($\mu$s)} &
  \multicolumn{1}{c}{Span} \\ 
  \cmidrule(lr){6-8}
  &
  \multicolumn{1}{c}{(ms)} &
  \multicolumn{1}{c}{($\textrm{cm}^{-3}$pc)} &
  \multicolumn{1}{c}{(d)} &
  \multicolumn{1}{c}{($\mu$s)} &
  med &
  $\mu$ &
  $\sigma$ &
  \multicolumn{1}{c}{(yr)} \\ \midrule

J0030+0451 & 4.87 & 4.33 & - & 62.38 & 3.11 & 3.54 & 2.18 & 1.80 \\
J0125-2327 & 3.68 & 9.59 & 7.28 & 66.04 & 0.71 & 0.99 & 0.95 & 2.43 \\
J0437-4715 & 5.76 & 2.65 & 5.74 & 87.51 & 0.13 & 0.13 & 0.01 & 2.50 \\
J0610-2100 & 3.86 & 60.69 & 0.29 & 87.86 & 2.77 & 3.04 & 1.61 & 2.43 \\
J0613-0200 & 3.06 & 38.79 & 1.20 & 38.09 & 1.01 & 1.25 & 0.83 & 2.43 \\
J0614-3329 & 3.15 & 37.05 & 53.58 & 31.29 & 2.43 & 3.15 & 2.92 & 2.43 \\
J0636-3044 & 3.95 & 15.46 & - & 106.92 & 3.91 & 5.32 & 4.26 & 2.37 \\
J0711-6830 & 5.49 & 18.41 & - & 79.99 & 3.39 & 4.67 & 4.15 & 2.44 \\
J0900-3144 & 11.11 & 75.69 & 18.74 & 224.47 & 2.17 & 2.36 & 0.88 & 2.44 \\
J0931-1902 & 4.64 & 41.49 & - & 57.20 & 3.37 & 4.17 & 2.97 & 2.34 \\
...        & ...   & ...   & ...   & ...  & ... & ... & ... & ... \\ \bottomrule
\end{tabular*}%
\end{table*}%

\subsection{Data Reduction and RFI mitigation}
\label{Subsection: data_reduction}

All MPTA observations were recorded using the PTUSE backend \citep{2020PASA...37...28B} and then transferred to both the SARAO data archive and the MeerTime data archive hosted on the OzStar supercomputer. The files were produced in \textsc{psrfits} \citep{2004PASA...21..302H} archive format, containing $8$-second sub-integrations, a time resolution of $1024$ bins per folding period, $4$ polarisation products (Stokes I, Q, U $\&$ V), and $1024$ frequency channels dividing the L-band digitized bandwidth ($856 - 1712$ MHz). All of the data are coherently dedispersed at the nominal dispersion measure of the
pulsar and folded at the topocentric period of the pulsar.
Computational limitations with the initial PTUSE machines meant we were not able to record the entire band for earlier observations, instead, we recorded the inner $928$ channels with a total bandwidth of $775.75$ MHz. This has since been remediated, however later data was reduced to keep a consistent band. The system temperature increases in the outer channels so is a relatively small loss in total sensitivity.
Changes in delays and time tagging within the hardware have also introduced known offsets in the pulsar timing data sets. These time offsets have been directly corrected for in this release. The epoch and size of the offsets (often referred to as `jumps') are listed in Table \ref{table: jump_table}. 

The custom processing pipeline, \textsc{meerpipe}, was used to further process the archives to a desirable format, and to perform required radio frequency interference (RFI) excision. \textsc{meerpipe} calibrates the full resolution data products in both flux and polarisation. Prior to the $9^\mathrm{th}$ of April 2020, polarisation calibration was performed through the use of the \textsc{psrchive} \footnote{http://psrchive.sourceforge.net/} \citep{2004PASA...21..302H} \texttt{pac} utility. Any data produced after this date are calibrated on-site by the SARAO Science Data Processing pipeline using calibration solutions derived from phase up solutions \citep{2021MNRAS.505.4483S}, prior to its folding in the PTUSE machines. A right-handed circular polarisation convention was adopted, and the pipeline converted the polarisation parameters from the native polarisation (a description of polarisation through the intensity of the two linear components and their cross terms), to the more familiar Stokes parameters. Following this, RFI excision was performed via \textsc{meerguard}, a modified version of the RFI-excision program \textsc{coastguard} \citep{2016MNRAS.458..868L}, which has been optimized for the RFI expected to be present at the MeerKAT telescope location. This software operates by comparing observations of a pulsar to a high-quality, frequency-averaged template profile. The residuals created from subtracting the observation from the profile are then used to identify channels where RFI is present, and the channels are subsequently zero-weighted in the archive. Lastly, the data were decimated to a more suitable format for use in this data release. 
The format of these files were frequency averaged from the original $928$ frequency channels, to $16$ sub-bands across the usable frequency range, and were fully time integrated over the entire observation length. These operations are both performed through the \texttt{pam} program contained within \textsc{psrchive}. The nature of the MPTA observing strategy results in most data output by \textsc{meerpipe} to have an integrated length of $\sim 256$ seconds.

\section{The MPTA Data Set}
\label{Section: data set / data products}

The MPTA data release is comprised of the full data archives, fully integrated pulse profiles, the portraits used for the creation of ToAs, the timing ephemerides used in this work, as well as sub-banded and band-averaged ToAs, for all pulsars listed in Table \ref{table: MPTA_summary_full_table}.

\begin{figure*}
    
    \includegraphics[width=\textwidth]{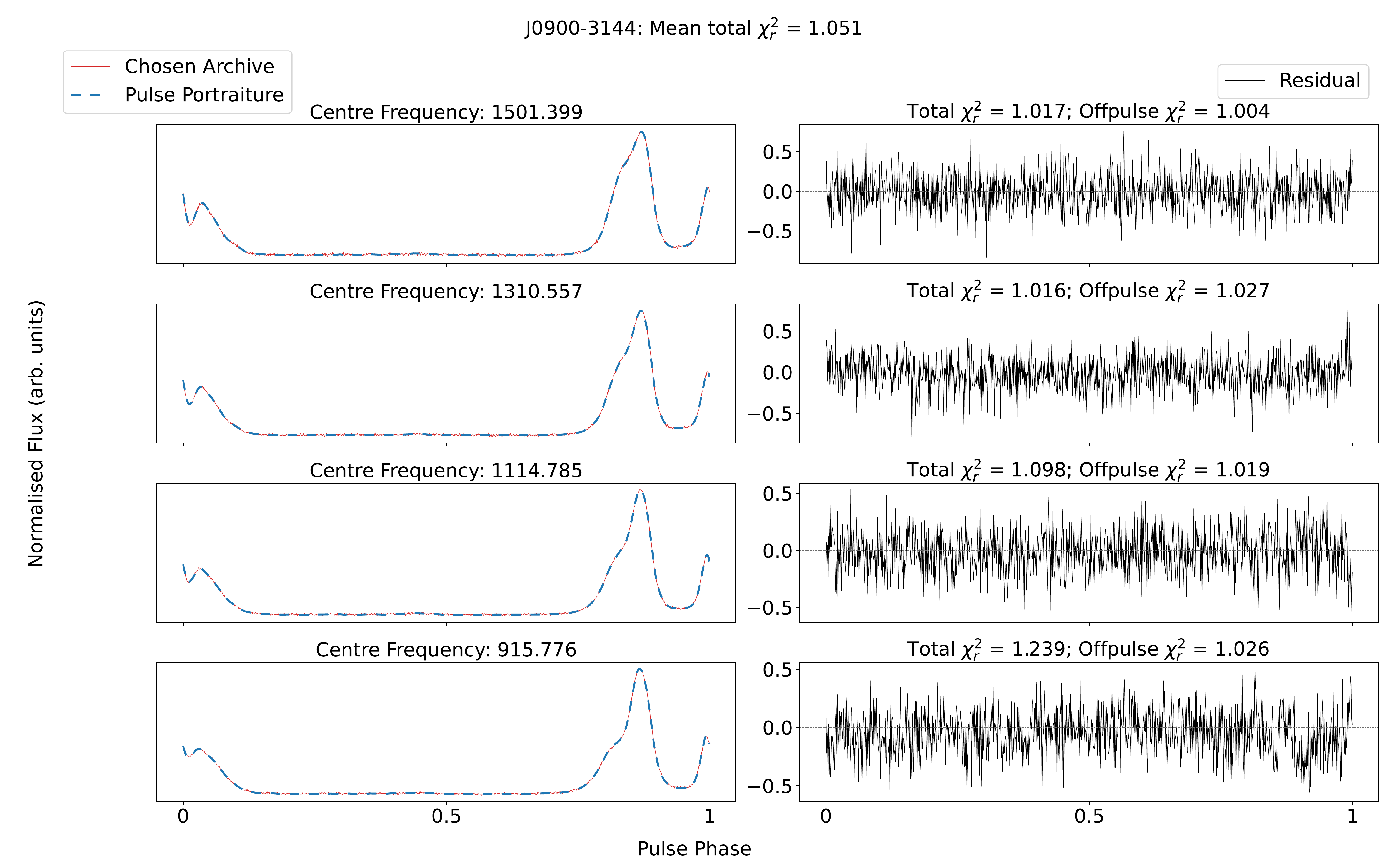}
    \vspace{-0.5cm}
    
    \includegraphics[width=\textwidth]{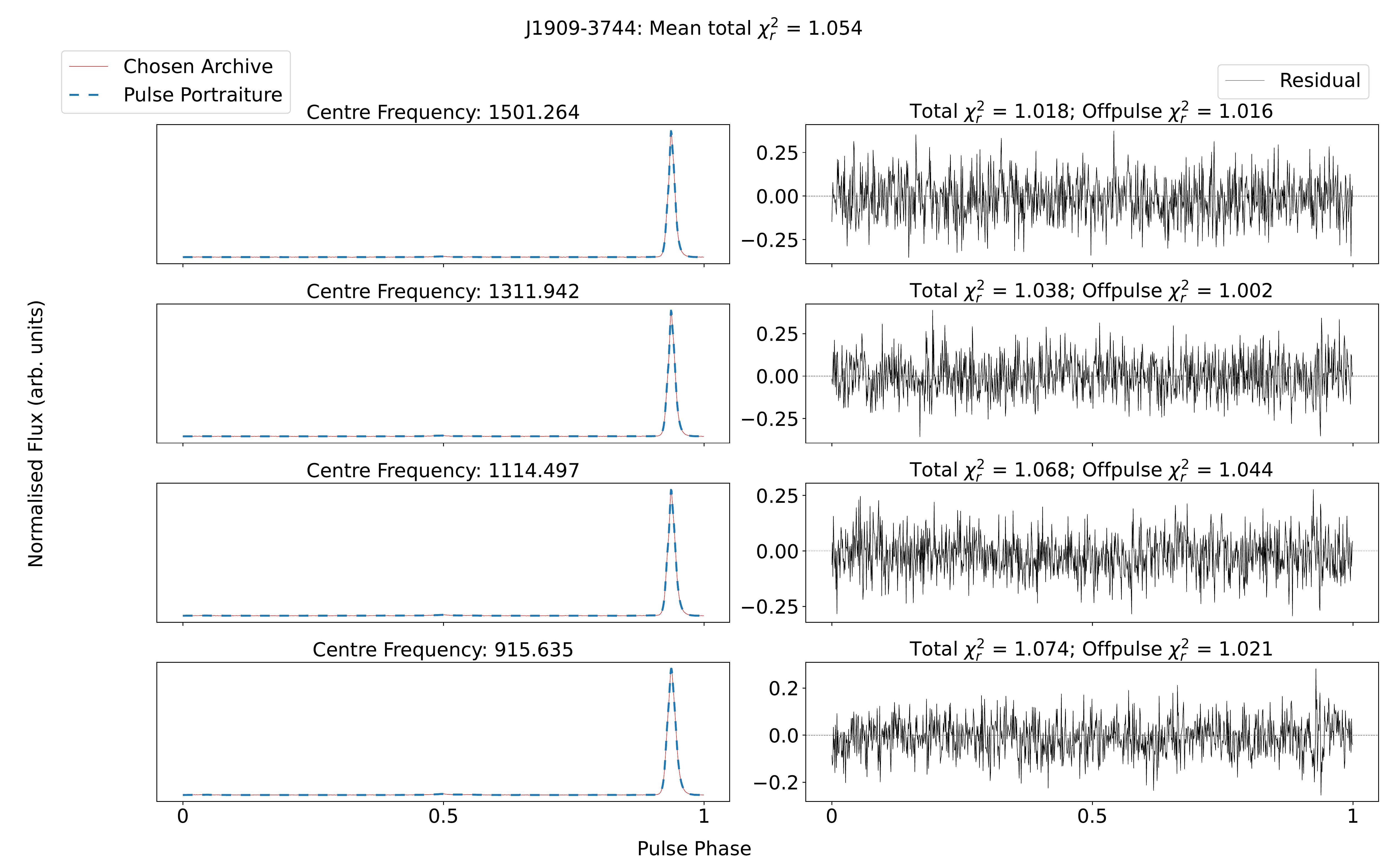}
    \caption[]{Comparison of MPTA portraits built from the full MPTA 2.5 year dataset, via the process described in Section \ref{Subsection: Portrait_creation}, with time-averaged observational data for J0900$-$3144 (Top) and J1909$-$3744 (Bottom). (Left) Direct comparison of \textsc{PulsePortraiture} models (dashed blue) against the observational data (solid red). (Right) The profile residuals from subtracting the model from the observational data. Above each residual, the reduced chi-squared statistic is supplied for both the total pulse profile, and for the off-pulse region. The portraits match the data well enough that it is difficult to discern the differences between the model and the data without the assistance of the residual plot.}
    \label{fig: portrait_example}
\end{figure*}

\subsection{Arrival times}
\label{Subsection: arrival_times}

To produce the arrival times we used the following procedure in the spirit of creating a reproduceable data set.

\begin{enumerate}
    \item From the $16$ sub-banded data output by \textsc{meerpipe}, identify and remove any observations with issues known to be unrecoverable. For clarity, an observation was defined as unrecoverable where there existed phase smearing of $>1$ bin in either the time or frequency space, or where the observation was taken with an incorrect position beyond the tied beam.
    \item Compensate for the system induced, \textit{a priori} known time delays, as per Table \ref{table: jump_table}.
    \item Form initial ToAs from a standard \textsc{psrchive} \texttt{pat} utility, using a Fourier domain monte-carlo algorithm (FDM), and output a tempo2 IPTA format.
    \item Filter for sub-banded observations with signal-to-noise ratio (S/N) $> 10$.
    \item Use \textsc{tempo2} \citep{2006MNRAS.369..655H} to fit for pulsar spin frequency and the frequency time derivative, as well as dispersion measure and its first time derivative. For some pulsars, these initial ephemerides were available from existing PTA data releases \cite[e.g.,][]{2021MNRAS.507.2137R}. For others, we started with ephemerides from the ATNF pulsar catalogue \citep{2005AJ....129.1993M}. Some parameters were of insufficient precision, resulting in significant phase drifts in some of the pulsar timing residuals. To account for this, additional timing parameters were selectively fit for in five pulsars. These parameters included right ascension, declination, orbital parameters, and proper motion. 
    \item Inspect outliers and record any that were removed, including a justification for exclusion.
    \item Create an intermediate ephemeris and timing files from \textsc{tempo2}.
    \item Update the data with the ephemeris, and create a corresponding new standard template.
    \item Create a new set of ToAs excluding the outliers. The ToAs were produced as in step (iii).
\end{enumerate}
Following these steps, the pulse profiles were further used to create portraits as described below, and sub-banded ToAs were subsequently calculated using the FDM algorithm and the filtered and updated data. The ephemerides included in this data release are preliminary as we have not fully explored the timing models or undertaken a comprehensive noise analysis. Nevertheless, they are suitable for gravitational wave and other parameter searches.

\subsection{Portrait Creation}
\label{Subsection: Portrait_creation}

Profile evolution, where pulsar-emission characteristics change with frequency, results in the need for frequency dependent average pulse profiles.  This is especially true with the high-sensitivity wide-band observations taken with MeerKAT. We detected profile evolution (of varying degree) in all  pulsars presented in the data set. The portraits used for timing were made with the \textsc{PulsePortraiture} software \citep{2019ApJ...871...34P}, which creates noise-free, high-quality, frequency-dependent templates, that can be used for precise sub-band timing. Models for two pulsars that have been created this way are shown in Figure \ref{fig: portrait_example}. 

\begin{figure}
    
    \includegraphics[width=\columnwidth]{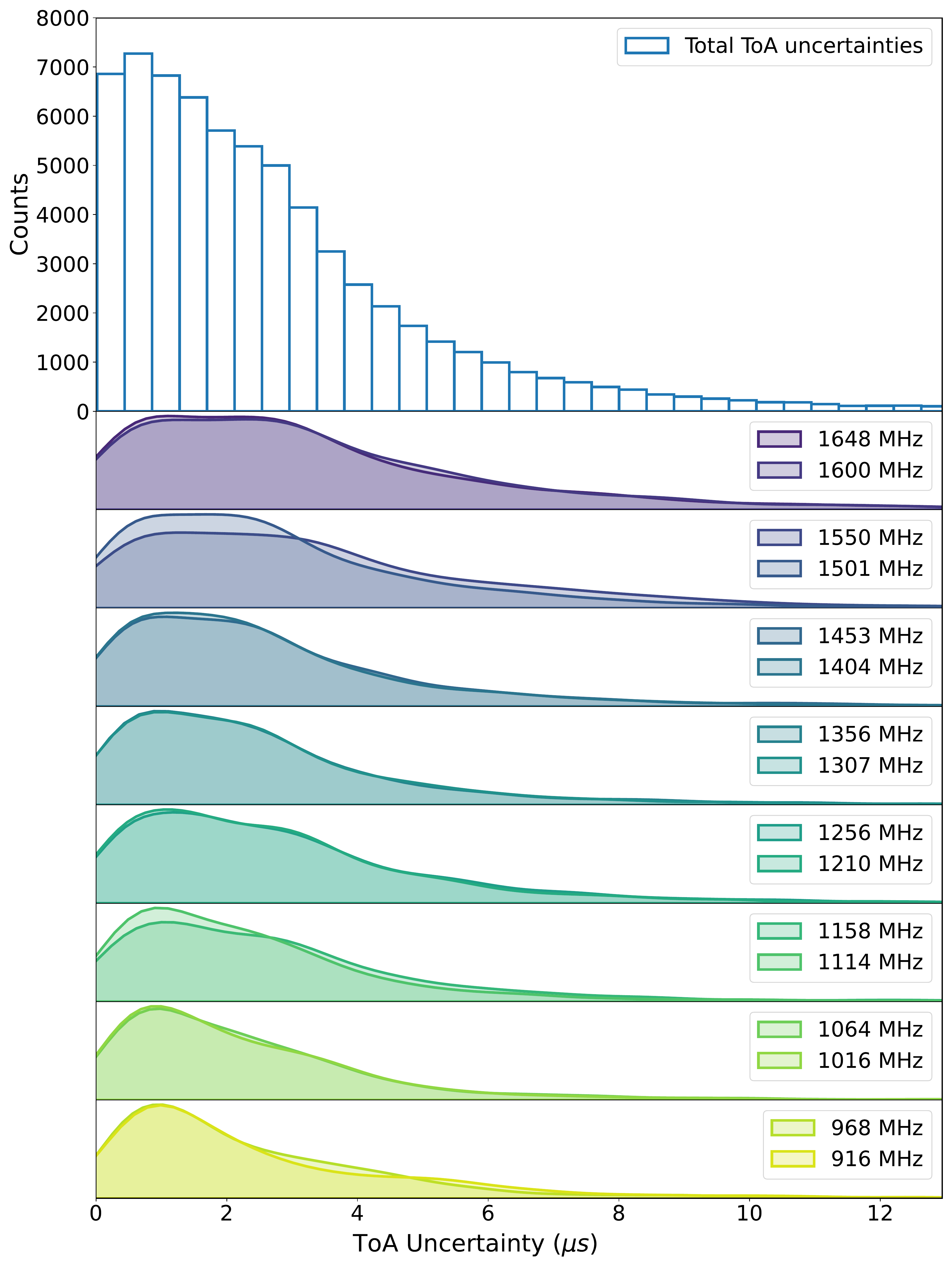}
    \caption[]{Arrival time distributions for the MPTA pulsars. The ToA uncertainties for all $78$ pulsars are included, the top panel showing the distribution of the uncertainties of ToAs from all sub-bands. The subsequent 8 panels show the density distribution of the ToAs that make up the total ToA uncertainties, separated by frequency, with the corresponding central frequency of each sub-band labelled. The largest $1\%$ of ToA uncertainties have been excluded from this plot for clarity, and can be treated as non-representative outliers.}
    \label{fig: toa_unc_w_subbands}
\end{figure}

Producing the pulse portraits was an iterative process. To form the pulse portraits, we first identify high S/N observations. The S/N threshold was varied between pulsars, as each pulsar possessed different median flux values, and many showed frequency dependent scintillation. The full frequency resolution ($928$ channel) archives from these observations were then combined using the \texttt{psradd} utility of \textsc{psrchive} using the pulsar ephemeris to align the observations. \textsc{PulsePortraiture} was used to decompose the profile using principal-component analysis (PCA), with the components allowed to vary in frequency independently.
We found that the process was sensitive to low levels of latent RFI in the data. This is because, in our use of \textsc{PulsePortraiture}, we normalize the data per frequency channel such that the profile evolution becomes the dominant source of variation as opposed to changes in intensity\footnote{We note there are multiple ways to scale the data in \textsc{PulsePortraiture}. However, this method is standard practice in pulsar timing \citep{2019ApJ...871...34P}.}. This normalisation amplified signals in nearly zero-weighted channels affected by RFI. To mitigate this we used \texttt{paz} to median filter, both before and after normalisation. Any remaining channels were manually excised using the \textsc{psrchive} program, \texttt{pazi}. Approximately $150$ channels were removed during this process, which is consistent with the channels that are affected by the RFI as reported in \citet{2020PASA...37...28B}. This was necessary as \textsc{PulsePortraiture} was sensitive to the presence of RFI. Similarly, while the construction of the frequency-dependent model returns a high level of agreement, demonstrated through the small uncertainty distributions in Figure \ref{fig: toa_unc_w_subbands}, elements of structure were still seen to exist in the profile residuals (the difference between the portraits and the data used to produce them), particularly in the lowest and highest sub-bands.
One possible reason for this may be that there exists latent RFI in the archives, that is not significant enough to have been automatically or manually removed, but may influence the components decomposed from the PCA. Alternatively, it could be that there exist subtle variations in the average pulse that were not well modelled through \textsc{PulsePortraiture}. Due to this, in the process of creating the timing residuals, frequency-dependant (FD) parameters \citep{2015ApJ...813...65N} were added to the majority of the pulsar parameter files (discussed in Section \ref{Subsection: Noise modelling}), with the aim to compensate for unmodelled frequency dependent variations. In addition, we note that in the creation of a frequency-resolved portrait, a fiducial DM must be chosen. This could result in differing frequency-resolved portraits, for different choices of DM. When combining MeerKAT data with other IPTA data sets, this may result in both PTA dependent FD parameters and DMs. Alternatively, one could attempt to re-time all the data sets with a common portrait.

\subsection{Noise Modelling and interpretation}
\label{Subsection: Noise modelling}

Our noise modelling followed approaches taken by other PTAs \citep{2017ApJ...834...35L, 2021MNRAS.502..478G, 2022MNRAS.509.5538C}. We modelled the influence of systematic noise that is uncorrelated through time with the noise parameters EFAC ($\mathrm{E}_\mathrm{{F}}$) and EQUAD ($\mathrm{E}_\mathrm{{Q}}$). $\mathrm{E}_\mathrm{{F}}$ is a parameter used directly as a scale factor of the timing uncertainties associated with the pulsar, and $\mathrm{E}_\mathrm{{Q}}$ influences the timing uncertainties as an addition in quadrature, such that the final timing uncertainties follow the convention \citep{2014MNRAS.437.3004L},

\begin{equation}
\label{eq: white_noise_BM}
    \sigma = \sqrt{\mathrm{E}^{2}_\mathrm{Q} + \mathrm{E}^{2}_\mathrm{F} \times \sigma_\mathrm{ToA}},
\end{equation}

\noindent where $\sigma_\mathrm{ToA}$ are the uncertainties reported by the time-tagging algorithm. We note that the convention of Equation \ref{eq: white_noise_BM} is  different to that used in \textsc{tempo2}, but consistent with that used in \textsc{temponest} \cite[][]{2014MNRAS.437...21M}. For the purposes of our model selection, we refer to this as our \textit{basic model}, or $\mathrm{M}_\mathrm{1}$ noise. In addition to this, another white noise parameter, ECORR ($\mathrm{E}_\mathrm{{C}}$), is included \citep{2015ApJ...813...65N}. This parameter accounts for noise that is correlated between sub-banded ToAs, but uncorrelated between observations. While statistically motivated, this component could account for stochastic pulse-pulse changes in pulse morphology, colloquially referred to as pulse jitter \citep{2014MNRAS.443.1463S, 2019ApJ...872..193L, 2021MNRAS.502..407P}. 

In addition to the three white-noise terms, we also searched for evidence of time-correlated noise processes that could be present in the MPTA pulsars. In particular, we modelled frequency-dependent DM variations, and an achromatic red-noise process. Both were modelled as stationary, stochastic processes with temporal variations of the form

\begin{equation}
\label{eq: power_spec_red}
    \mathrm{S}(f) \propto \frac{\mathrm{A}^{2}}{12 \pi^{2}} \left( \frac{f}{f_{c}} \right)^{-\gamma},
\end{equation}

\noindent where A is the signal amplitude, $\gamma$ is the spectral index of the power spectrum, and $f_{\mathrm{c}} =1\,\mathrm{yr}^{-1}$ is the reference frequency. 

Prior to determining noise models for each pulsar, a rudimentary significance test was performed on whether frequency-dependent parameters (FD1/FD2), and dispersion measure derivatives (DM1/DM2) were required in the parameter files. These parameters were determined to be significant, and therefore required to be fitted for, if it was found that the value of the parameter was $>3 \sigma$.

To search for the  possible noise sources in each pulsar, we considered eight models created from different possible combinations of the noise processes described here, shown in Table \ref{table: noise_models}. We used the \textsc{temponest} \citep{2014MNRAS.437.3004L} software to perform a Bayesian analysis to appropriately model the noise, whilst also analytically marginalising over the timing model \cite[see Appendix A of ][]{2009MNRAS.395.1005V}). The preferred model was selected by comparing the relative evidences for each model using the following procedure. We first considered the model with the largest evidence. If it was found to have a log-Bayes factor, $\rm{\ln}\mathcal{B}$, of $\rm{\ln}\mathcal{B} \geq 4$ over all other competing models, we selected it as the preferred model. If this was not the case, a model that possessed fewer noise terms that fulfilled the requirement of $\rm{\ln}\mathcal{B} \geq 4$ over other competing models was chosen. Where there were no models that satisfied this requirement, the basic model ($\mathrm{M}_{1}$) was chosen.

Following this model selection process, $31$ of the pulsars present in this data set show evidence for time-correlated DM noise. Significant DM variations in other MSP data sets have been shown \citep{2007MNRAS.378..493Y} to occur on observation spans of equivalent lengths to the data we present here. As such, the presence of time-correlated DM variations is likely to be found at the $\sim 2.5$ yr scale of this release, especially given the precision of the data provided. Generally, the turbulent ISM that is thought to give rise to DM variations in timing data is modelled through a Kolmogorov spectrum \citep{1990ARA&A..28..561R, 2021MNRAS.502..478G} and, assuming this, we would expect a $\gamma_\mathrm{DM} \sim 8/3$. This is relatively inconsistent with our findings, with the majority ($20$) of the pulsars that show evidence for DM noise having spectral indices that are less steep (Table \ref{table: MPTA_results_full}). However, in direct comparison to previous noise studies, e.g. \citet{2021MNRAS.502..478G}, we found that DM noise in many of the MPTA pulsars also found in the PPTA are comparable to within reported uncertainties.

We found evidence for achromatic red noise in $4$ pulsars (Table \ref{table: MPTA_results_full}). The presence of this noise source is of note as it can significantly impact the sensitivity of the pulsar to gravitational waves, particularly if the noise has similar spectral indices to what is expected of a gravitational wave background. Aside from the SGWB, there are a number of possibe sources of red noise, including pulsar intrinsic noise such as interactions between the superfluidic centre of a neutron star and its crust \citep{2014MNRAS.437...21M}, magnetospheric state switching \citep{2010Sci...329..408L}, or small bodies orbiting the pulsar \citep{2013ApJ...766....5S}. 
Additionally we note that the values of $\mathrm{E}_\mathrm{{C}}$ obtained both in this data release and elsewhere are historically larger than what is seen in independent studies of noise due to pulse jitter \citep{2021MNRAS.502..407P}, and so it is possible that $\mathrm{E}_\mathrm{{C}}$ includes contributions from another stochastic noise source that has not yet been fully characterised.
We also note that for some pulsars the reduced-$\chi^2$ is less than unity. In many cases, we suspect that this is due to the influence of the $\mathrm{E}_\mathrm{{C}}$ noise parameter presenting at a greater amplitude than expected. An alternative explanation in pulsars possessing fewer observing epochs and binary orbital parameters, is that the generalised least squares fit performed by \textsc{tempo2} is absorbing some of the noise in the deterministic timing parameters.


\begin{table}

\caption{Models compared to determine the most probable representation of the noise processes in each pulsar.}
\label{table: noise_models}
\resizebox{\columnwidth}{!}{%
\begin{tabular}{@{}ccccc@{}}
\toprule
Model & $\mathrm{E}_\mathrm{Q}$ + $\mathrm{E}_\mathrm{F}$ & $\mathrm{E}_\mathrm{C}$ & DM Noise & \begin{tabular}[c]{@{}c@{}}Achromatic \\ Red Noise\end{tabular} \\ \midrule
$\mathrm{M}_{1}$ & \checkmark & -          & -          & -          \\
$\mathrm{M}_{2}$ & \checkmark & \checkmark & -          & -          \\
$\mathrm{M}_{3}$ & \checkmark & -          & -          & \checkmark \\
$\mathrm{M}_{4}$ & \checkmark & -          & \checkmark & -          \\
$\mathrm{M}_{5}$ & \checkmark & \checkmark & -          & \checkmark \\
$\mathrm{M}_{6}$ & \checkmark & \checkmark & \checkmark & -          \\
$\mathrm{M}_{7}$ & \checkmark & \checkmark & \checkmark & \checkmark \\
$\mathrm{M}_{8}$ & \checkmark & -          & \checkmark & \checkmark \\ \bottomrule
\end{tabular}%
}
\end{table}

\subsection{Residuals and noise subtraction}
\label{Subsection: residuals}

\begin{figure}
    
    \includegraphics[width=\columnwidth]{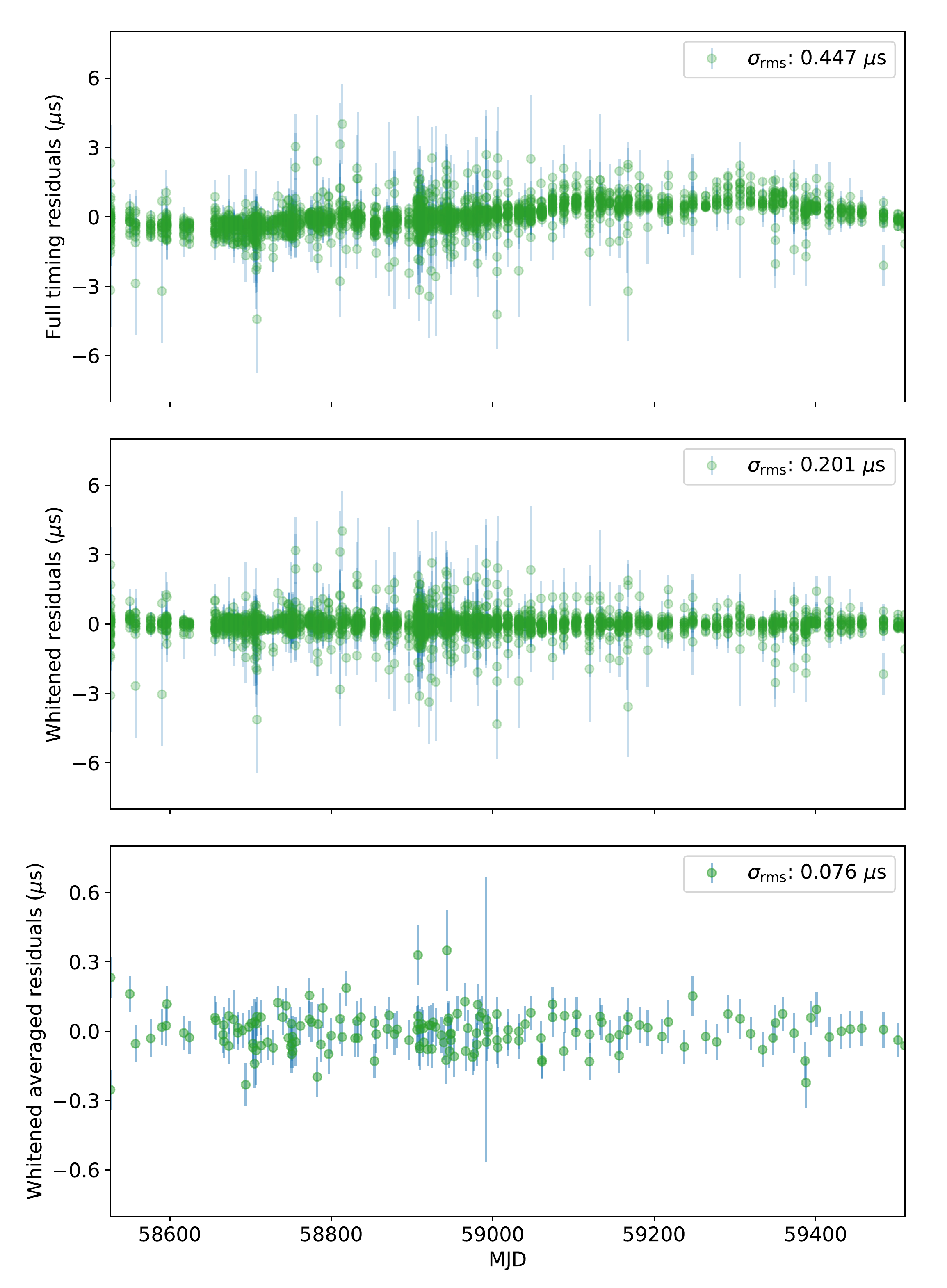}
    \caption[]{Timing residuals for PSR J1909$-$3744. (Top) Full sub-banded timing residuals with no noise reduction. (Middle) Sub-banded timing residuals fully whitened via the removal of all identified red-noise processes. (Bottom) Fully whitened band-averaged timing residuals. Note the change in y-axis scale by a factor of 10 for ease of contrast.}
    \label{fig: 1909_noise_comparison}
\end{figure}

We used the timing residuals to assess the quality of the data. Two sets of timing residuals were formed from the sub-banded data: the first with frequency resolution where each sub-banded ToA correlates directly to a timing residual, and the second in which the timing residuals have been averaged together such that a single timing residual is produced for all ToAs that are recorded simultaneously. It was clear through visual inspection that some pulsars possessed evidence for time correlated noise, even on the 2.5-year data release timeline.
We can measure the impact of the noise processes on the data by comparing the residuals before and after subtracting the maximum likelihood realisation of the process. An example of this can be seen in Figure \ref{fig: 1909_noise_comparison}.

In Table \ref{table: MPTA_results_short} we present the MPTA timing results, and the comparisons between the models we have described here for ten pulsars. The full results are provided in Table \ref{table: MPTA_results_full}.


\begin{table*}

\caption{MPTA timing results and indicative noise parameters for the first ten MPTA pulsars in order of Right Ascension. The full results of all timing analyses, and the full indicative noise parameters, can be found in Table \ref{table: MPTA_results_full}.}
\label{table: MPTA_results_short}
\makebox[\textwidth]{%
\begin{tabular}[c]{lrccccccccccc}
\toprule
Pulsar &
  \multicolumn{1}{l|}{$\textrm{N}_{\textrm{ToA}}$} &
  \multicolumn{1}{c|}{$\mathrm{M}_\mathrm{N}$} &
  \multicolumn{4}{c|}{Time-Correlated Noise} &
  \multicolumn{2}{c|}{\begin{tabular}[c]{@{}c@{}}Sub-banded \\ wRMS ($\mu$s)\end{tabular}} &
  \multicolumn{4}{c}{\begin{tabular}[c]{@{}c@{}}Band-averaged \\ wRMS ($\mu$s)\end{tabular}} \\
  \cmidrule(lr){4-7}\cmidrule(lr){8-9}\cmidrule(lr){10-13}
   &
  \multicolumn{1}{l|}{} &
  \multicolumn{1}{c|}{} &
  $\mathrm{log}_{10}\textrm{A}_{\textrm{DM}}$ &
  \multicolumn{1}{c|}{$\gamma_{\textrm{DM}}$} &
  $\mathrm{log}_{10}\textrm{A}_{\text{red}}$ &
  $\gamma_{\textrm{red}}$ &
  Full &
  Whitened &
  Full &
  $\chi^{2}_{\mathrm{r},\mathrm{full}}$ &
  Whitened &
  $\chi^{2}_{\mathrm{r},\mathrm{white}}$ \\ \midrule

J0030+0451 & 747 & 1 & - & - & - & - & 1.895 & - & 0.457 & 1.146 & - & - \\
J0125-2327 & 893 & 6 & -12.3(1) & 4.0(2) & - & - & 0.718 & 0.634 & 0.549 & 2.114 & 0.402 & 1.130 \\
J0437-4715 & 1052 & 6 & -11.8(1) & 1.3(4) & - & - & 0.298 & 0.184 & 0.268 & 2.513 & 0.157 & 0.859 \\
J0610-2100 & 801 & 4 & -11.4(2) & 1.4(8) & - & - & 1.947 & 1.838 & 0.658 & 2.762 & 0.295 & 0.555 \\
J0613-0200 & 687 & 1 & - & - & - & - & 0.832 & - & 0.220 & 1.804 & - & - \\
J0614-3329 & 700 & 1 & - & - & - & - & 1.482 & - & 0.367 & 1.106 & - & - \\
J0636-3044 & 540 & 1 & - & - & - & - & 2.095 & - & 0.411 & 0.697 & - & - \\
J0711-6830 & 775 & 1 & - & - & - & - & 1.475 & - & 0.431 & 1.436 & - & - \\
J0900-3144 & 896 & 6 & -10.9(1) & 1.8(5) & - & - & 3.587 & 2.328 & 2.624 & 5.308 & 0.985 & 0.748 \\
J0931-1902 & 419 & 1 & - & - & - & - & 1.856 & - & 0.602 & 1.679 & - & - \\ 
... & ... & ... & ... & ... & ... & ... & ... & ... & ... & ... & ... & ... \\ \bottomrule
\end{tabular}%
}
\end{table*}%

\section{Discussion}
\label{Section: Discussion}
\subsection{Findings from the MPTA Data Set}
\label{Subsection: Data set performance}

The initial timing analysis demonstrates the potential of the MPTA. Of the pulsars included in this data release $26$ are unique to the MPTA, and $67$ are able to achieve sub-microsecond precision using the current observing strategy. In comparison to the major PTAs, the MPTA leads in this performance statistic. Band-averaged, whitened sub-microsecond precision is achieved for $13$ of the PPTA pulsars \citep{2020PASA...37...20K}, for $5$ of the EPTA pulsars \citep{2016MNRAS.458.3341D}, and for $44$ of the pulsars regularly monitored by NANOGrav \citep{2021ApJS..252....4A}.

In the coming years, the primary goal of the MPTA will be the detection of nanohertz-frequency gravitational radiation, collaborating with the IPTA towards the goal of the detection of an angularly correlated signal. It is important to note that while the sensitivity of a PTA to the SGWB increases strongly with the length of the data span, it is also directly proportional to the number of pulsars in the array \citep{2013CQGra..30v4015S}. As noted in \citet{spiewak2022}, the inclusion of the MPTA and its otherwise unobserved pulsars, results in the MPTA potentially becoming the most significant contributor to IPTA searches for the SGWB by 2025, unless cadences or sensitivities at the other PTAs change. In addition, there is work currently underway investigating improvements to the MPTA, potentially resulting in a more aggressive timeline for improved sensitivity (Middleton et al., in prep.). 

The sensitivity of the MPTA data set enabled investigations of the relationship between the change in the profile shape through frequency, and the expected timing precision. As noted in Section \ref{Subsection: Portrait_creation}, at least some degree of profile evolution through frequency was noted in all the pulsars in this data release. The effective pulse width,
\begin{equation}
\label{eq: eff_width}
    \mathrm{W_{eff}} = \frac{\Delta \phi}{\Sigma_{\mathrm{i}}[\mathrm{P}(\phi_{\mathrm{i}+1}) - \mathrm{P}(\phi_{\mathrm{i}})]^{2}},
\end{equation}
where $\Delta \phi$ is the resolution of the profile in pulse phase, $\phi$ is defined as the positional pulse phase, and P is the pulse profile flux normalized to unity, is measured for each pulsar's frequency sub-bands. By weighting each sub-band observation by its corresponding S/N, a linear relationship between the $\textrm{W}_\textrm{eff}$ and the timing uncertainty is empirically derived across the MPTA dataset,
\begin{equation}
\label{eq: uncertainty_width}
    \sigma_{\textrm{s}}(\textrm{W}_\textrm{eff}) = \alpha \frac{\textrm{W}_\textrm{eff}}{(\textrm{S}/\textrm{N})_\textrm{obs,s}},
\end{equation}
where $\sigma_\textrm{s}$ is the ToA uncertainty at each sub-band, and $(\textrm{S}/\textrm{N})_\textrm{obs,s}$ is the S/N for each sub-band of the observation. We use the \texttt{polyfit} module of the \textsc{numpy} python package to measure the coefficient of this relationship to be $\alpha = 1.249 \pm 0.002$ and the relative scatter on this approximation is defined as,
\begin{equation}
\label{eq: var_unc_width}
    \delta = \frac{1}{N}\sum\frac{|\sigma_\mathrm{obs} - \sigma_\mathrm{s}(\textrm{W}_\textrm{eff})|}{\sigma_\mathrm{s}(\textrm{W}_\textrm{eff})},
\end{equation}
where N is the number of data points and $\sigma_\mathrm{obs}$ is the uncertainty at each observation. Evaluating this expression results in $\delta = 0.201$, implying a fractional uncertainty of $\sim 20\%$ of the data relating to this model.

We also investigated how the effective width varies with frequency. Naively, one would expect the effective width to vary monotonically with frequency. For example, if the pulse profile is affected by scattering from a thin screen in the ISM we would expect the relationship between the thin-screen scattering timescale ($\tau_\textrm{s}$) and frequency ($f$) to evolve as \citep{1977ARA&A..15..479R},

\begin{equation}
\label{eq: scattering_timescale}
    \tau_\textrm{s} \propto f^{-4}.
\end{equation}

It is also thought that higher-frequency emission mechanisms occur closer to the surface of the pulsar, resulting in narrower pulse widths from the relative radius of the emission cone \citep{1978ApJ...222.1006C, 1999ApJ...526..957K, 2015MNRAS.449.3223D}.

\begin{figure}
    
    \includegraphics[width=\columnwidth]{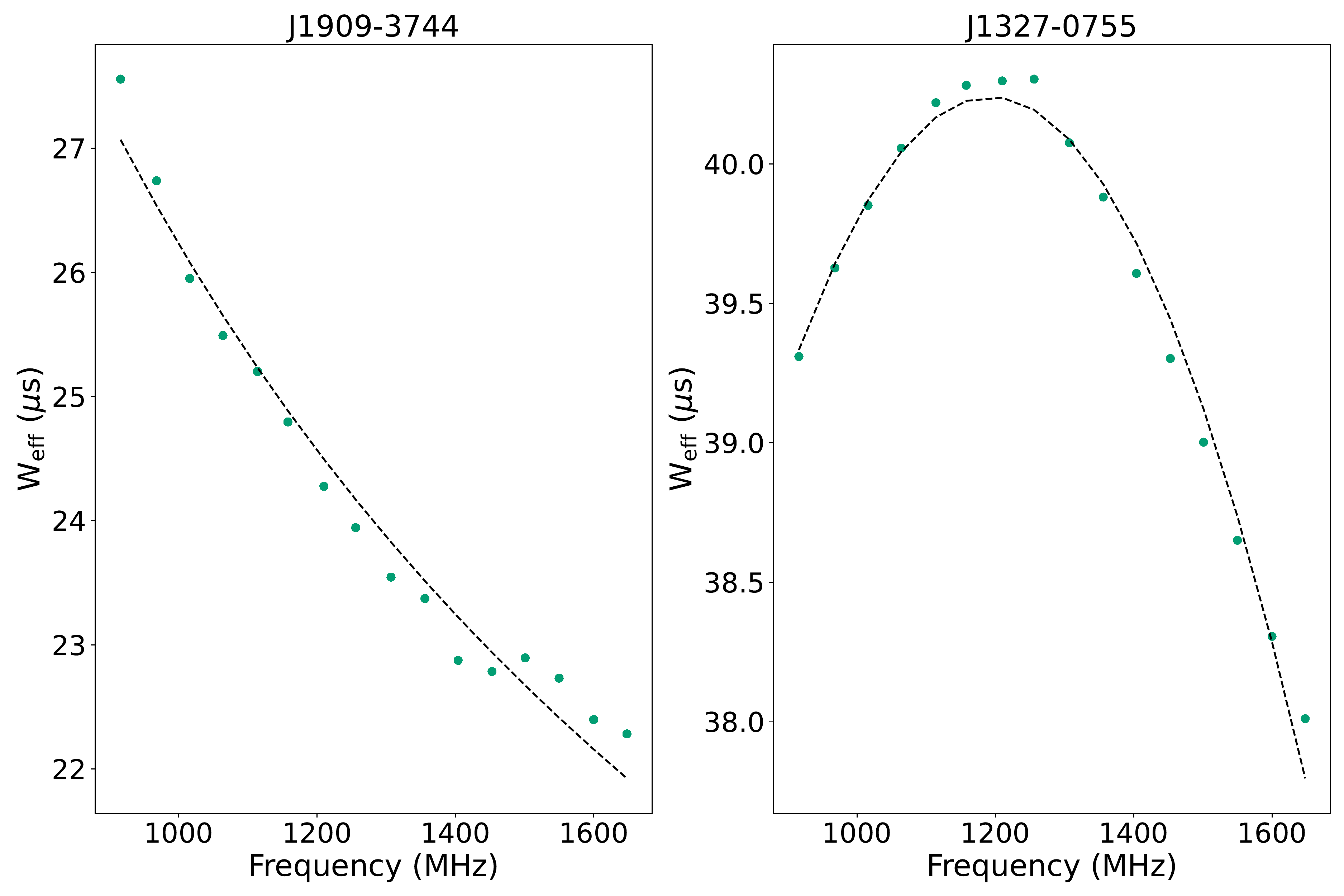}
    \caption[]{Two pulsars (Left: J1909-3744; Right: J1327-0755) that possess decidedly different relationships between effective width ($W_\text{eff}$) and frequency. The relationship exhibited by J1909$-$3744 is seen more often, whereas J1327$-$0755 exhibits nonmonotonic, and  almost parabolic, evolution through frequency.}
    \label{fig: effective_width_unc}
\end{figure}

While we found $44$ of the MPTA pulsars follow power-law relationships of effective width and frequency, others demonstrated far more complicated relationships better described by polynomials of order $>1$ (Figure \ref{fig: effective_width_unc}). These relationships are counter-intuitive to how pulsar emission is thought to occur, as well as how it is thought to be affected by the ISM, with recent studies of non-recycled pulsars identifying an expected, largely monotonic, width evolution through frequency with both positive and negative slopes \citep{2021MNRAS.508.4249P}. By combining the effective width measurements with the flux density it will be possible to further optimise observing strategies, and assess the utility of observations in other bands at MeerKAT, or with other telescopes such as the SKA.

\subsection{Clock Correction Recovery}
\label{Subsection: Clock recovery}

The main purpose of a PTA is to detect signals common to the ensemble of pulsars. To demonstrate the potential of the MPTA, we use the data release to recover the MeerKAT observatory time standard.

The MeerKAT station clock uses two Hydrogen masers, two Rubidium atomic clocks, and a quartz crystal. This station clock produces a time standard (Karoo Telescope Time, KTT). The time standard drifts by a few nanoseconds per day relative to the Coordinate Universal Time (UTC) as reported by the Global Positioning Service (GPS). Every few months the masers are adjusted to keep the time within 500~ns of UTC (GPS) \citep{2020PASA...37...28B}. 

For accurate pulsar timing, the drifting offset of KTT relative to UTC  is measured through a system described in detail in \citet{2022JATIS...8a1013B}, and results in the production of a clock correction\footnote{https://archive-gw-1.kat.ac.za/public/tfr/mk2utc.clk}. This correction is applied using pulsar timing software (in our case \textsc{tempo2}), which corrects for the resultant offset in the pulse arrival times to a precision of less than $4.925$ ns \citep{2022JATIS...8a1013B}. 

Here we derive the KTT from the pulsar timing measurements. To do this, we use the approach described in \citet{2020MNRAS.491.5951H}, in order to constrain the properties of the clock signal through a Bayesian methodology. This was chosen as opposed to the frequentist methods, which is also presented in \citet{2020MNRAS.491.5951H}, as the Bayesian approach is now adopted by most PTA collaborations.

We modelled the clock signal as a power-law, red noise process of the form
\begin{equation}
\label{eq: clock_power_spectrum}
    S_{\mathrm{clk}}(f) = \frac{{A^{2}_{\mathrm{clk}}}}{12\pi^{2}}\left(\frac{f}{f_c}\right)^{-\gamma_\mathrm{clk}},
\end{equation}
where the coefficients are of the same form as Equation \ref{eq: power_spec_red}. 

To perform parameter estimation we employed a Parallel Tempered Markov Chain Monte Carlo (PTMCMC) \citep{justin_ellis_2017_1037579} sampler via the \textsc{enterprise} software \citep{2019ascl.soft12015E}, and allowed \textsc{enterprise} to simultaneously model individual pulsar red noise processes (isolated to a general achromatic red noise process and a dispersion measure noise process) as well as a common red noise process of the form given in Equation \ref{eq: clock_power_spectrum} across all pulsars in the sample. This parameter estimation resulted in a common signal possessing the noise characteristics $\log_{10}\mathrm{A}_\mathrm{clk} = -13.35$ and $\gamma_\mathrm{clk} = 2.1$.

Following this, we constructed noise covariance matrices for each pulsar of the form
\begin{equation}
\label{eq: cov_total}
    \textbf{C}_\mathrm{PSR} = \textbf{C}_\mathrm{white} + \textbf{C}_\mathrm{red} + \textbf{C}_\mathrm{DM},
\end{equation}
where $\textbf{C}_\mathrm{white}$ models the white noise components, including $\mathrm{E}_\mathrm{C}$ (noise that is correlated between sub-banded ToAs) where required. Without the addition of $\mathrm{E}_\mathrm{C}$, this is a diagonal matrix populated with the arrival time uncertainties, influenced by the white noise terms as per Equation \ref{eq: white_noise_BM}. With the addition of $\mathrm{E}_\mathrm{C}$, the matrix is no longer diagonal, but accounts for a covariance value between arrival times from the same observation. 

$\textbf{C}_\mathrm{DM}$ and $\textbf{C}_\mathrm{red}$ are the covariance matrices relating to the DM noise and the achromatic red noise processes, accounting for their presence where required. We followed the definitions of the covariance matrices presented in \citet{2014MNRAS.441.2831L}. The covariance between the ToAs $i$ and $j$ are

\begin{equation}
\label{eq: covariance_DM}
    \textbf{C}_\mathrm{DM,ij} = \frac{\kappa^{2}\int_{1/T}^{\infty}\mathrm{S}_\mathrm{DM}(f)\cos{(2\pi f t_{ij})}\,df}{\nu^{2}_\mathrm{i} \nu^{2}_\mathrm{j}},
\end{equation}
and
\begin{equation}
\label{eq: covariance_red}
    \textbf{C}_\mathrm{red,ij} = \int_{1/T}^{\infty}\mathrm{S}_\mathrm{red}(f)\cos{(2\pi f t_{ij})}\,df, 
\end{equation}
where T is the total time span of the data, $f$ is the fluctuation frequency (as opposed to observing frequency $\nu$), $\mathrm{t}_\mathrm{ij}$ is the time lag between arrival times i and j, and $\kappa$ is the dispersion constant $\kappa = 2.410 \times 10^{-16}$ $\mathrm{Hz}^{-2}\mathrm{cm}^{-3}\mathrm{pc}\mathrm{s}^{-1}$. 
$\mathrm{S}_\mathrm{red}(f)$ and $\mathrm{S}_\mathrm{DM}(f)$ are the power spectral densities of the achromatic red noise and DM noise respectively and are defined as,
\begin{equation}
\label{eq: red_power_spectrum}
    S_{\mathrm{red}}(f) = \frac{{A^{2}_{\mathrm{red}}}}{12\pi^{2}}\left(\frac{f}{f_c}\right)^{-\gamma_\mathrm{red}},
\end{equation}
and
\begin{equation}
\label{eq: DM_power_spectrum}
    S_{\mathrm{DM}}(f) = {A^{2}_\mathrm{DM}}\left(\frac{f}{f_c}\right)^{-\gamma_\mathrm{DM}},
\end{equation}
noting the lack of the $1/12\pi^{2}$ normalisation factor in Equation \ref{eq: DM_power_spectrum}, following the convention of \citet{2014MNRAS.437.3004L}.

Using this methodology we could reconstruct the clock signal over an arbitrary date range, that can extend beyond our data span. 
We did this by the use of ten of the most precise pulsars presented in this data release, combining the covariance matrices we describe above via
\begin{equation}
\label{eq: mega_matrix}
    \textbf{C}_\mathrm{PTA,10} = \begin{bmatrix}
\textbf{C}_{1} & 0 & 0 & ... & 0 \\
0 & \textbf{C}_{2} & 0 & ... & 0 \\
0 & 0 & \textbf{C}_{3} & ... & 0 \\
... & ... & ... & ... & ... \\
0 & 0 & 0 & ... & \textbf{C}_{10} 
\end{bmatrix},
\end{equation}

where $\textbf{C}_\mathrm{N}$ represents each of the 10 pulsars' respective $\textbf{C}_\mathrm{PSR}$ per Equation \ref{eq: cov_total}. 

In order to evaluate the clock signal, we defined $N_\mathrm{rec}=500$ epochs at which we wanted to reconstruct it. We chose the initial epoch to be when KTT was first published at MJD 58484, and we chose to extend the reconstruction beyond our observations to MJD 59611, to investigate the ability to extrapolate the clock signal.

To do this we then extended $\textbf{C}_\mathrm{PTA,10}$ as per \citet{2012MNRAS.424..244D}, by $N_\mathrm{rec}$ rows and columns to create a matrix of the form

\begin{equation}
\label{eq: final_matrix}
    \textbf{C}_\mathrm{Final} = \begin{bmatrix}
\textbf{C}_\mathrm{PTA,10} & 0 \\
0 &  0 
\end{bmatrix},
\end{equation}
populated by zeroes.

The clock covariance is then evaluated at not only our observations, but at the epochs of our reconstructed signal

\begin{equation}
    \textbf{C}_\mathrm{Total} = \textbf{C}_\mathrm{Final} + \textbf{C}_\mathrm{clk},
\end{equation}

where the covariance of the clock signal is defined as
\begin{equation}
\label{eq: covariance_clock}
    \textbf{C}_\mathrm{clk,ij} = \int_{1/T_\mathrm{clk}}^{\infty}\mathrm{S}_\mathrm{clk}(f)\cos{(2\pi f t_{ij})}\,df,
\end{equation}
with equation parameters as described in Equations \ref{eq: covariance_DM} and \ref{eq: covariance_red}. We note that the clock covariance is calculated simultaneously for both the epochs of the observations, and the epochs for the reconstructed signal. Further, $T_\mathrm{clk}$ is the span over which the clock signal is evaluated.

We defined the total residuals $\textbf{t}_\mathrm{res}$ at which this signal is evaluated, as the vector containing the observed timing residuals for the 10 pulsars $\textbf{t}_\mathrm{pf,PTA} = (\textbf{t}_\mathrm{pf,1}, ..., \textbf{t}_\mathrm{pf,10})$, and the perfect residuals we assigned to the epochs of our reconstructed signal $\textbf{t}_\mathrm{ideal}=\textbf{0}$.

This final covariance matrix was used to measure the recovered clock waveform \citep{2014MNRAS.441.2831L, 2020MNRAS.491.5951H},

\begin{equation}
\label{eq: clock_waveform}
    \textbf{t}_\mathrm{clk} = \textbf{C}_\mathrm{clk}\textbf{C}^{-1}_\mathrm{Total}\textbf{t}_\mathrm{res}.
\end{equation}

The covariance of the clock signal is
\begin{equation}
\label{eq: clock_uncertainties}
    \mathbf{\sigma}_\mathrm{clk} = \textbf{C}_\mathrm{clk} - \textbf{C}_\mathrm{clk}\textbf{C}^{-1}_\mathrm{Total}\textbf{C}_\mathrm{clk},
\end{equation}
where the uncertainty of the clock signal is the square root of the diagonal of $\mathbf{\sigma}_\mathrm{clk}$.

\begin{figure*}
    
    \includegraphics[width=\textwidth]{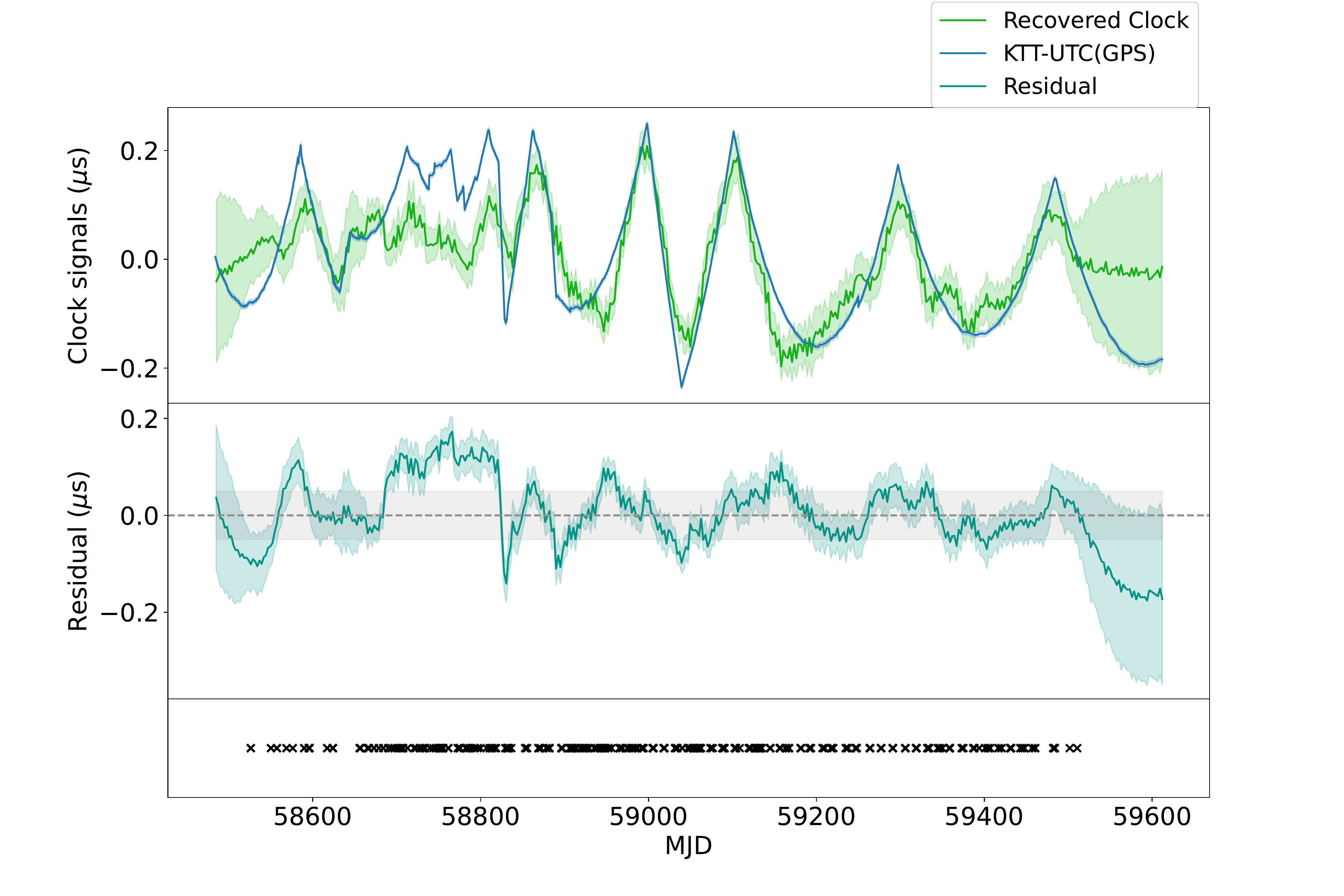}
    \caption[]{(Top) The recovered clock correction signal and corresponding uncertainty as defined in Section \ref{Subsection: Clock recovery} (green, shaded), juxtaposed against the clock correction signal that was to be extracted from the data and corresponding uncertainty (blue, shaded). Note the uncertainty rapidly increasing following the end of the data span, as well as the marked deviation between MJDs $\sim58700$ and $\sim58900$. (Middle) The residual difference (teal) between the recovered clock signal and the KTT-UTC(GPS) clock correction signal. The teal shading corresponds to the uncertainties of both the recovered signal and KTT-UTC(GPS) summed together in quadrature. The dashed grey line and corresponding shading represent a $50$ ns window around $0$ ns, to demonstrate the precision of the recovered signal. (Bottom) The MJDs corresponding to observations used for this reconstruction.}
    \label{fig: clock_recovery}
\end{figure*}

The recovered clock signal is presented in Figure \ref{fig: clock_recovery}. The precision to which the MPTA is able to extract an injected signal is evident, recovering the signal to within $50$ ns for the majority of the data span. There is evidence of a $\approx 150$\,n s deviation in the signal between MJDs $\sim58700$ and $\sim58900$, the cause of which is currently uncertain.
While the beginning of this apparent  offset is close to one of the known timing offsets in the data, the end is not associated with one.  

First, we considered if the deviation was caused by an individual pulsar with poorly specified noise. To test this, we removed one of the pulsars from the method and reformed the clock signal. We found the feature persisted in all cases, suggesting an effect common to all of the pulsars. Possible other causes of the deviation could be unidentified systematic errors in the clock correction, an additional noise process that is common to the pulsars, or even the influence of a SGWB. 

Owing to the unknown pulsar spin frequency and frequency derivatives, pulsars are not sensitive to linear or quadratic variations in time standards. We attempted to account for a possible quadratic drift in KTT, and found that the residual of the expected and the recovered clock signal is not able to be well modelled by a quadratic. We conclude that the deviation we observe is not expected to be caused by a quadratic. We also note that the signal we reconstructed is not well-described by a Gaussian process. Arising from this misspecification, it is possible that the reconstructed clock signal will not well recover the sharp (high fluctuation frequency) features in KTT. This also may contribute to the larger discrepancy between MJDs $\sim58700$ and $\sim58900$. 

Uncertainties that exist in a telescope time standard must be kept far lower than the amplitude of the expected SGWB signal. Typically, observatory time standards are assumed to have precision much lower than the GW signal, however, this is the first time that this has been demonstrated through pulsar timing \citep{2016MNRAS.455.4339T}. The difference between the recovered clock corrections and the corresponding time standard can then be used as a measure of the accuracy of a timing array. We propose that the root-mean-square (rms) residual and its spectral characteristics form a quantitative measure of the quality of the timing array. These can then be compared with the amplitudes of any purported gravitational wave background to assess array performance. For the MPTA the rms residual is 62 ns over 2.5 years. Assuming a SGWB amplitude of $2 \times 10^{-15}$ (similar to the amplitude observed in the purported common red signal), the expected induced rms variations in five years of observational data is expected to be to be $\sim 127$ ns \cite[][]{2013MNRAS.428.1147V}.  

\section{Summary and Conclusions}
\label{Section: Summary and Conclusions}

We have presented here the first data release from the MPTA, including the observations of $78$ pulsars spanning February 2019 to October 2021. As well as the recorded sub-banded and band-averaged arrival times for each pulsar, we also provide initial timing ephemerides, timing portraits, sub-banded and band-averaged timing residuals, and preliminary noise models. Of these pulsars, $67$ report sub-microsecond timing precision, with $2$ possessing a precision of less than $100$ ns from whitened, band-averaged data. We detail the observing strategy of the MPTA, and selection criteria for a pulsar to be deemed suitable for inclusion into it. Additionally, the data reduction pipeline used for this data release, and the nature of the data products output by it is outlined. We also describe how the accompanying portraits are created through the use of \textsc{PulsePortraiture}, and the process of noise model selection and the consequent subtraction of noise models from the data in order to provide more accurate timing residuals. All data products described here are open to the scientific community, and can be accessed via the Australian gravitational wave data portal\footnote{http://dx.doi.org/10.26185/6392814b27073}. The collaboration expects to release subsequent data releases with both the four and five year MPTA data sets when they are mature.

We have also investigated the relationship of the effective-width metric ($\mathrm{W}_\mathrm{eff}$) and observing frequency, and find somewhat counter-intuitive results that may influence the observing priority in frequency bands outside of the L-band range we report on. We provide initial noise models, showing that DM variations are detected in a substantial sub-sample of our pulsars. These noise models can be used as the starting point to develop more sophisticated models.

Lastly, we are able to demonstrate the precision of the MPTA data set by using a subset of the full MPTA to recover the observatory time standard, Karoo Telescope Time (KTT). Through most of the data we see good agreement, with the recovered signal within $50$ ns of KTT. We find a modest ($150$ ns) disagreement for $\sim 150$ days that appears to be common to all the pulsars. Work is ongoing to see if we can explain this discrepancy. We test if this deviation could be attributed to any individual pulsar we include in the signal recovery.  As we finding that it cannot, we briefly explore possible other causes.

The MPTA data set includes the largest number of regularly timed pulsars of any current PTA. In this data release, we do not include an analysis of the merits of wide-band timing methods as opposed to the sub-banded arrival times we calculate, and would encourage any interested parties to fully utilise the components of this data release to explore this technique. Individually, the data released here has the potential for a wide range of studies, many of which are already well underway by members of the MeerTime collaboration. This includes the study of individual pulsars including updated measurements of orbital parameters. We can also study the population as a whole. This includes an analysis of the astrometry in MPTA pulsars, aiming to improve MSP population-synthesis models; an investigation of the flux density variations in the MPTA population, both to study the ISM through diffractive and refractive scintillation, and to optimise pulsar timing observations via dynamic scheduling; and, naturally, searches for common noise and correlated signals to identify the influence of gravitational waves on the MPTA pulsars, as well as other signals that are common in the array.

Finally, the inclusion of the MPTA data set to the IPTA data streams is expected to make a significant impact in the coming years \citep{spiewak2022}, both through the inclusion of high-quality timing data of pulsars that are already observed, and through the inclusion of data from an additional 26 pulsars that are not yet observed in any other PTA.

\section*{Acknowledgements}

We thank T. Pennucci for discussions. 
The MeerKAT telescope is operated by the South African Radio Astronomy Observatory, which is a facility of the National Research Foundation, an agency of the Department of Science and Innovation.
MTM, RMS, MB, and DR acknowledge support through the ARC centre of Excellence grant CE17010004 (OzGrav). 
RMS acknowledges support through Australian Council Future Fellowship FT190100155.
VVK acknowledges continuing support from the Max Planck society.
GT acknowledges financial support from Agence Nationale de la Recherche (ANR-18-CE31-0015), France.
AK acknowledges funding from the STFC consolidated grant to Oxford Astrophysics, code ST/000488.
PTUSE was developed with support from the Australian SKA Office and Swinburne University of Technology, with financial contributions from the MeerTime collaboration members.
This work used the OzSTAR national facility at Swinburne University of Technology. 
OzSTAR is funded by Swinburne University of Technology and the National Collaborative Research Infrastructure Strategy (NCRIS).

\section*{Data Availability}

All data used in this release available at the Australian Gravitational Wave Data Centre, at \url{http://dx.doi.org/10.26185/6392814b27073}. 
The data provided includes both sub-banded and frequency averaged ToAs, the full data archives used to construct this data release, the ephemerides that have been used to perform timing, and band-averaged and sub-banded timing residuals. Also included are both the timing standards (frequency averaged, one dimensional) and frequency resolved portraits, used to calculate each set of ToAs as well as the preliminary noise models constructed to facilitate this work.

The archives, standards, and portraits are in \textsc{psrfits} file format. The ephemerides and residuals are in a standard \texttt{ascii} text file format, and the arrival times are supplied as IFF data. Information regarding the preliminary noise models can be found contained within the ephemerides.
Data from four pulsars (PSRs J0614$-$3329, J1036$-$8317, J1756$-$2251, and J1902$-$5105) will be available after the 18-month SARAO embargo is lifted. An updated DOI will be provided at this time.



\bibliographystyle{mnras}
\bibliography{main}




\appendix
\label{Appendix}
\section{MPTA timing results and indicative noise parameters}

\clearpage
\onecolumn
\setlength\LTleft{0pt}
\setlength\LTright{0pt}
\begin{longtable}[c]{@{}lrccccccccccc@{}}
\captionsetup{width=0.9\textwidth}
\caption{MPTA timing results and indicative noise characteristics. Included here for each pulsar are the number of sub-banded ToAs included in the data release ($\textrm{N}_{\textrm{ToA}}$), the noise model used as per Table \ref{table: noise_models} ($\mathrm{M}_\mathrm{N}$), the noise characteristics of the pulsar, and the precision of the pulsar's timing residuals, both before and after subtracting the maximum likelihood realisation of the pulsar's noise processes. We provide this for in terms of the weighted root-mean-square uncertainty (wRMS) for both the sub-banded timing residuals and the band-averaged timing residuals. For the band-averaged timing residuals, we provide a reduced chi-squared statistic for comparison of the impact of removing the influence of the noise processes. We note here that the whitened residuals remove all time-correlated noise processes associated with the pulsar.}
\label{table: MPTA_results_full}\\
\toprule
Pulsar &
  \multicolumn{1}{l}{$\textrm{N}_{\textrm{ToA}}$} &
  \multicolumn{1}{c}{$\mathrm{M}_\mathrm{N}$} &
  \multicolumn{4}{c}{Time-Correlated Noise} &
  \multicolumn{2}{c}{\begin{tabular}[c]{@{}c@{}}Sub-banded \\ wRMS ($\mu$s)\end{tabular}} &
  \multicolumn{4}{c}{\begin{tabular}[c]{@{}c@{}}Band-averaged \\ wRMS ($\mu$s)\end{tabular}} \\
  \cmidrule(lr){4-7}\cmidrule(lr){8-9}\cmidrule(lr){10-13}
\endfirsthead
\multicolumn{13}{c}%
{{\bfseries Table \thetable\ continued from previous page}} \\
\toprule
Pulsar &
  \multicolumn{1}{l}{$\textrm{N}_{\textrm{ToA}}$} &
  \multicolumn{1}{c}{$\mathrm{M}_\mathrm{N}$} &
  \multicolumn{4}{c}{Time-Correlated Noise} &
  \multicolumn{2}{c}{\begin{tabular}[c]{@{}c@{}}Sub-banded \\ wRMS ($\mu$s)\end{tabular}} &
  \multicolumn{4}{c}{\begin{tabular}[c]{@{}c@{}}Band-averaged \\ wRMS ($\mu$s)\end{tabular}} \\
  \cmidrule(lr){4-7}\cmidrule(lr){8-9}\cmidrule(lr){10-13}
   &
  \multicolumn{1}{l}{} &
  \multicolumn{1}{c}{} &
  $\mathrm{log}_{10}\textrm{A}_{\textrm{DM}}$ &
  \multicolumn{1}{c}{$\gamma_{\textrm{DM}}$} &
  $\mathrm{log}_{10}\textrm{A}_{\text{red}}$ &
  $\gamma_{\textrm{red}}$ &
  Full &
  Whitened &
  Full &
  $\chi^{2}_{\mathrm{r},\mathrm{full}}$ &
  Whitened &
  $\chi^{2}_{\mathrm{r},\mathrm{white}}$ \\* \midrule
\endhead
\bottomrule
\endfoot
\endlastfoot
 &
  \multicolumn{1}{l}{} &
  \multicolumn{1}{c}{} &
  $\mathrm{log}_{10}\textrm{A}_{\textrm{DM}}$ &
  \multicolumn{1}{c}{$\gamma_{\textrm{DM}}$} &
  $\mathrm{log}_{10}\textrm{A}_{\text{red}}$ &
  $\gamma_{\textrm{red}}$ &
  Full &
  Whitened &
  Full &
  $\chi^{2}_{\mathrm{r},\mathrm{full}}$ &
  Whitened &
  $\chi^{2}_{\mathrm{r},\mathrm{white}}$ \\* \midrule
J0030+0451 & 747 & 1 & - & - & - & - & 1.895 & - & 0.457 & 1.146 & - & - \\
J0125-2327 & 893 & 6 & -12.3(1) & 4.0(2) & - & - & 0.718 & 0.634 & 0.549 & 2.114 & 0.402 & 1.130 \\
J0437-4715 & 1052 & 6 & -11.8(1) & 1.3(4) & - & - & 0.298 & 0.184 & 0.268 & 2.513 & 0.157 & 0.859 \\
J0610-2100 & 801 & 4 & -11.4(2) & 1.4(8) & - & - & 1.947 & 1.838 & 0.658 & 2.762 & 0.295 & 0.555 \\
J0613-0200 & 687 & 1 & - & - & - & - & 0.832 & - & 0.220 & 1.804 & - & - \\
J0614-3329 & 700 & 1 & - & - & - & - & 1.482 & - & 0.367 & 1.106 & - & - \\
J0636-3044 & 540 & 1 & - & - & - & - & 2.095 & - & 0.411 & 0.697 & - & - \\
J0711-6830 & 775 & 1 & - & - & - & - & 1.475 & - & 0.431 & 1.436 & - & - \\
J0900-3144 & 896 & 6 & -10.9(1) & 1.8(5) & - & - & 3.587 & 2.328 & 2.624 & 5.308 & 0.985 & 0.748 \\
J0931-1902 & 419 & 1 & - & - & - & - & 1.856 & - & 0.602 & 1.679 & - & - \\
J0955-6150 & 2238 & 6 & -10.9(1) & 2.7(7) & - & - & 3.224 & 2.545 & 2.425 & 3.723 & 1.260 & 1.005 \\
J1012-4235 & 892 & 1 & - & - & - & - & 4.756 & - & 0.958 & 0.795 & - & - \\
J1017-7156 & 1056 & 6 & -11.3(1) & 1.4(3) & - & - & 0.774 & 0.405 & 0.790 & 9.487 & 0.245 & 0.912 \\
J1022+1001 & 748 & 6 & -11.3(5) & 2.0(2) & - & - & 2.273 & 2.066 & 1.575 & 1.493 & 1.240 & 0.925 \\
J1024-0719 & 734 & 1 & - & - & - & - & 1.232 & - & 0.297 & 1.124 & - & - \\
J1036-8317 & 649 & 1 & - & - & - & - & 1.336 & - & 0.388 & 1.413 & - & - \\
J1045-4509 & 848 & 4 & -10.9(1) & 1.7(4) & - & - & 3.489 & 2.381 & 2.451 & 23.367 & 0.464 & 0.853 \\
J1101-6424 & 956 & 4 & -11.0(1) & 2.9(9) & - & - & 3.836 & 3.389 & 1.812 & 6.319 & 0.786 & 1.086 \\
J1103-5403 & 813 & 2 & - & - & - & - & 3.334 & - & 3.712 & 1.016 & - & - \\
J1125-5825 & 845 & 6 & -11.4(2) & 5.0(1) & - & - & 3.153 & 2.138 & 2.405 & 8.692 & 0.844 & 1.062 \\
J1125-6014 & 1007 & 6 & -11.7(2) & 4.0(1) & - & - & 0.791 & 0.420 & 0.671 & 19.729 & 0.147 & 0.948 \\
J1216-6410 & 880 & 4 & -11.7(2) & 1.6(6) & - & - & 0.967 & 0.913 & 0.334 & 2.880 & 0.162 & 0.672 \\
J1327-0755 & 208 & 1 & - & - & - & - & 1.066 & - & 0.381 & 1.132 & - & - \\
J1421-4409 & 833 & 1 & - & - & - & - & 3.816 & - & 0.973 & 1.332 & - & - \\
J1431-5740 & 846 & 4 & -10.6(1) & 2.3(5) & - & - & 5.692 & 3.742 & 4.057 & 25.707 & 0.671 & 0.676 \\
J1435-6100 & 955 & 2 & - & - & - & - & 2.506 & - & 1.646 & 1.068 & - & - \\
J1446-4701 & 639 & 4 & -11.5(3) & 3.0(1) & - & - & 1.632 & 1.317 & 0.959 & 8.733 & 0.332 & 1.036 \\
J1455-3330 & 660 & 1 & - & - & - & - & 1.679 & - & 0.485 & 1.866 & - & - \\
J1525-5545 & 905 & 6 & -10.6(1) & 2.0(1) & - & - & 6.508 & 3.947 & 5.539 & 3.967 & 2.747 & 0.975 \\
J1543-5149 & 752 & 2 & - & - & - & - & 3.019 & - & 1.464 & 0.994 & - & - \\
J1545-4550 & 940 & 4 & -11.7(3) & 4.0(1) & - & - & 0.729 & 0.629 & 0.383 & 9.853 & 0.178 & 1.894 \\
J1547-5709 & 732 & 4 & -11.2(2) & 1.0(6) & - & - & 3.366 & 3.147 & 1.259 & 3.438 & 0.623 & 0.951 \\
J1600-3053 & 784 & 6 & -11.8(3) & 3.0(2) & - & - & 0.597 & 0.507 & 0.359 & 2.950 & 0.203 & 0.941 \\
J1603-7202 & 829 & 2 & - & - & - & - & 1.918 & - & 0.799 & 0.964 & - & - \\
J1614-2230 & 779 & 1 & - & - & - & - & 0.992 & - & 0.315 & 2.338 & - & - \\
J1629-6902 & 859 & 1 & - & - & - & - & 1.338 & - & 0.330 & 1.129 & - & - \\
J1643-1224 & 848 & 7 & -10.7(1) & 3.5(7) & -12.4(3) & 3.0(1) & 3.050 & 1.295 & 2.582 & 21.351 & 0.473 & 0.718 \\
J1652-4838 & 846 & 6 & -10.7(1) & 1.6(5) & - & - & 3.613 & 2.472 & 2.778 & 4.259 & 1.168 & 0.753 \\
J1653-2054 & 607 & 4 & -10.9(1) & 0.8(4) & - & - & 4.209 & 2.973 & 2.838 & 22.237 & 0.393 & 0.427 \\
J1658-5324 & 302 & 1 & - & - & - & - & 1.260 & - & 0.407 & 3.000 & - & - \\
J1705-1903 & 778 & 6 & -11.1(1) & 0.7(4) & - & - & 1.159 & 0.982 & 1.184 & 2.126 & 0.896 & 1.218 \\
J1708-3506 & 638 & 3 & - & - & -12.3(2) & 1.9(8) & 3.281 & - & 1.521 & 6.535 & - & - \\
J1713+0747 & 640 & 1 & - & - & - & - & 0.694 & - & 0.148 & 1.074 & - & - \\
J1719-1438 & 804 & 1 & - & - & - & - & 2.436 & - & 0.600 & 1.203 & - & - \\
J1721-2457 & 368 & 1 & - & - & - & - & 3.656 & - & 0.809 & 1.179 & - & - \\
J1730-2304 & 830 & 2 & - & - & - & - & 1.324 & - & 0.588 & 1.051 & - & - \\
J1732-5049 & 1007 & 2 & - & - & - & - & 1.334 & - & 0.857 & 1.051 & - & - \\
J1737-0811 & 943 & 4 & -11.1(1) & 0.8(5) & - & - & 4.349 & 3.966 & 1.895 & 4.573 & 0.832 & 0.886 \\
J1744-1134 & 817 & 2 & - & - & - & - & 0.405 & - & 0.254 & 1.028 & - & - \\
J1747-4036 & 893 & 7 & -10.8(1) & 2.1(5) & -12.7(3) & 5.0(1) & 3.210 & 2.012 & 2.277 & 8.422 & 0.670 & 0.728 \\
J1751-2857 & 843 & 1 & - & - & - & - & 3.002 & - & 0.838 & 1.583 & - & - \\
J1756-2251 & 2016 & 6 & -10.9(1) & 0.4(4) & - & - & 7.885 & 7.386 & 5.302 & 0.715 & 5.033 & 0.644 \\
J1757-5322 & 1118 & 2 & - & - & - & - & 1.493 & - & 0.584 & 0.966 & - & - \\
J1801-1417 & 848 & 1 & - & - & - & - & 2.444 & - & 0.634 & 1.311 & - & - \\
J1802-2124 & 832 & 6 & -10.7(1) & 1.7(5) & - & - & 4.107 & 3.243 & 3.827 & 1.903 & 2.662 & 0.921 \\
J1811-2405 & 1199 & 6 & -11.3(1) & 1.5(5) & - & - & 1.013 & 0.736 & 0.733 & 2.830 & 0.391 & 0.801 \\
J1825-0319 & 832 & 4 & -11.0(1) & 2.2(7) & - & - & 3.602 & 3.161 & 1.678 & 5.661 & 0.618 & 0.775 \\
J1832-0836 & 485 & 4 & -11.4(2) & 2.0(1) & - & - & 1.169 & 0.892 & 0.742 & 14.191 & 0.234 & 1.369 \\
J1843-1113 & 824 & 4 & -11.2(1) & 3.2(7) & - & - & 1.834 & 1.166 & 1.365 & 26.667 & 0.260 & 0.912 \\
J1902-5105 & 894 & 2 & - & - & - & - & 1.238 & - & 0.891 & 1.128 & - & - \\
J1903-7051 & 1247 & 2 & - & - & - & - & 0.979 & - & 0.592 & 1.540 & - & - \\
J1909-3744 & 2837 & 6 & -11.72(9) & 2.6(4) & - & - & 0.447 & 0.201 & 0.392 & 25.455 & 0.076 & 0.965 \\
J1918-0642 & 860 & 1 & - & - & - & - & 1.282 & - & 0.326 & 1.395 & - & - \\
J1933-6211 & 1445 & 1 & - & - & - & - & 0.999 & - & 0.227 & 1.428 & - & - \\
J1946-5403 & 571 & 3 & - & - & -13.0(3) & 2.0(1) & 0.727 & - & 0.433 & 7.310 & - & - \\
J2010-1323 & 805 & 1 & - & - & - & - & 1.431 & - & 0.389 & 1.398 & - & - \\
J2039-3616 & 588 & 1 & - & - & - & - & 1.063 & - & 0.312 & 1.846 & - & - \\
J2124-3358 & 904 & 1 & - & - & - & - & 1.109 & - & 0.312 & 1.498 & - & - \\
J2129-5721 & 852 & 1 & - & - & - & - & 0.731 & - & 0.197 & 1.296 & - & - \\
J2145-0750 & 863 & 2 & - & - & - & - & 1.098 & - & 0.732 & 1.053 & - & - \\
J2150-0326 & 652 & 2 & - & - & - & - & 1.271 & - & 0.675 & 1.142 & - & - \\
J2222-0137 & 990 & 2 & - & - & - & - & 1.286 & - & 0.781 & 0.783 & - & - \\
J2229+2643 & 591 & 1 & - & - & - & - & 1.810 & - & 0.453 & 2.369 & - & - \\
J2234+0944 & 632 & 2 & - & - & - & - & 1.956 & - & 1.046 & 1.354 & - & - \\
J2241-5236 & 1003 & 2 & - & - & - & - & 0.111 & - & 0.049 & 0.631 & - & - \\
J2317+1439 & 539 & 1 & - & - & - & - & 1.600 & - & 0.475 & 1.361 & - & - \\
J2322+2057 & 424 & 1 & - & - & - & - & 1.725 & - & 0.446 & 2.074 & - & - \\
J2322-2650 & 694 & 1 & - & - & - & - & 1.403 & - & 0.401 & 2.043 & - & - \\

\end{longtable}
\twocolumn

\section{MPTA parameters and statistics}

\clearpage
\onecolumn
\setlength\LTleft{0pt}
\setlength\LTright{0pt}
\begin{longtable}[c]{@{\extracolsep{\fill}}lcccccccc@{}}
\captionsetup{width=0.9\textwidth}
\caption{MPTA partial parameters and timing statistics. The parameters that are included here are the pulsar's rotational period (P), associated dispersion measure (DM), orbital period where the pulsar has a companion ($\mathrm{P}_\mathrm{b}$), and the frequency-averaged effective width as per Equation \ref{eq: eff_width} ($\mathrm{W}_\mathrm{eff}$). Also included are statistics of the pulsar's sub-banded arrival time uncertainties over its entire data span, which are provided as the median (med), mean ($\mu$), and standard deviation ($\sigma$), and the corresponding observing span for each pulsar by the MPTA (Span).}
\label{table: MPTA_summary_full_table}\\
\toprule
Pulsar &
  \multicolumn{1}{c}{P} &
  \multicolumn{1}{c}{DM} &
  \multicolumn{1}{c}{$\textrm{P}_{\textrm{b}}$} &
  \multicolumn{1}{c}{$\textrm{W}_{\textrm{eff}}$} &
  \multicolumn{3}{c}{Uncertainty ($\mu$s)} &
  \multicolumn{1}{c}{Span} \\ 
  \cmidrule(lr){6-8}
\endfirsthead
\multicolumn{9}{c}%
{{\bfseries Table \thetable\ continued from previous page}} \\
\toprule
Pulsar &
  \multicolumn{1}{c}{P} &
  \multicolumn{1}{c}{DM} &
  \multicolumn{1}{c}{$\textrm{P}_{\textrm{b}}$} &
  \multicolumn{1}{c}{$\textrm{W}_{\textrm{eff}}$} &
  \multicolumn{3}{c}{Uncertainty ($\mu$s)} &
  \multicolumn{1}{c}{Span} \\ 
  \cmidrule(lr){6-8}
  &
  \multicolumn{1}{c}{(ms)} &
  \multicolumn{1}{c}{($\textrm{cm}^{-3}$pc)} &
  \multicolumn{1}{c}{(d)} &
  \multicolumn{1}{c}{($\mu$s)} &
  med. &
  $\mu$ &
  $\sigma$ &
  \multicolumn{1}{c}{(years)} \\* \midrule
\endhead
\bottomrule
\endfoot
\endlastfoot
&
  \multicolumn{1}{c}{(ms)} &
  \multicolumn{1}{c}{($\textrm{cm}^{-3}$~pc)} &
  \multicolumn{1}{c}{(d)} &
  \multicolumn{1}{c}{($\mu$s)} &
  med. &
  $\mu$ &
  $\sigma$ &
  \multicolumn{1}{c}{(yr)} \\* \midrule 
J0030+0451 & 4.87 & 4.33 & - & 62.38 & 3.11 & 3.54 & 2.18 & 1.80 \\
J0125-2327 & 3.68 & 9.59 & 7.28 & 66.04 & 0.71 & 0.99 & 0.95 & 2.43 \\
J0437-4715 & 5.76 & 2.65 & 5.74 & 87.51 & 0.13 & 0.13 & 0.01 & 2.50 \\
J0610-2100 & 3.86 & 60.69 & 0.29 & 87.86 & 2.77 & 3.04 & 1.61 & 2.43 \\
J0613-0200 & 3.06 & 38.79 & 1.20 & 38.09 & 1.01 & 1.25 & 0.83 & 2.43 \\
J0614-3329 & 3.15 & 37.05 & 53.58 & 31.29 & 2.43 & 3.15 & 2.92 & 2.43 \\
J0636-3044 & 3.95 & 15.46 & - & 106.92 & 3.91 & 5.32 & 4.26 & 2.37 \\
J0711-6830 & 5.49 & 18.41 & - & 79.99 & 3.39 & 4.67 & 4.15 & 2.44 \\
J0900-3144 & 11.11 & 75.69 & 18.74 & 224.47 & 2.17 & 2.36 & 0.88 & 2.44 \\
J0931-1902 & 4.64 & 41.49 & - & 57.20 & 3.37 & 4.17 & 2.97 & 2.34 \\
J0955-6150 & 2.00 & 160.90 & 24.58 & 82.05 & 3.01 & 3.50 & 1.99 & 2.50 \\
J1012-4235 & 3.10 & 71.65 & 37.97 & 63.90 & 4.67 & 5.28 & 2.95 & 2.43 \\
J1017-7156 & 2.34 & 94.22 & 6.51 & 33.76 & 0.36 & 0.41 & 0.27 & 2.53 \\
J1022+1001 & 16.45 & 10.25 & 7.81 & 145.24 & 1.48 & 2.69 & 3.45 & 2.15 \\
J1024-0719 & 5.16 & 6.48 & - & 69.54 & 2.70 & 3.60 & 3.14 & 2.43 \\
J1036-8317 & 3.41 & 27.10 & 0.34 & 37.45 & 2.17 & 2.30 & 1.18 & 2.24 \\
J1045-4509 & 7.47 & 58.11 & 4.08 & 277.01 & 2.62 & 3.09 & 1.76 & 2.48 \\
J1101-6424 & 5.11 & 207.35 & 9.61 & 128.62 & 3.87 & 4.22 & 1.80 & 2.48 \\
J1103-5403 & 3.39 & 103.91 & - & 76.30 & 3.03 & 3.49 & 1.90 & 2.44 \\
J1125-5825 & 3.10 & 124.81 & 76.40 & 52.31 & 2.31 & 2.40 & 0.81 & 2.24 \\
J1125-6014 & 2.63 & 52.93 & 8.75 & 16.19 & 0.50 & 0.52 & 0.29 & 2.44 \\
J1216-6410 & 3.54 & 47.39 & 4.04 & 22.95 & 1.07 & 1.11 & 0.37 & 2.53 \\
J1327-0755 & 2.68 & 27.91 & 8.44 & 39.55 & 2.20 & 2.30 & 1.12 & 1.72 \\
J1421-4409 & 6.39 & 54.64 & 30.75 & 124.81 & 4.48 & 5.22 & 2.84 & 2.48 \\
J1431-5740 & 4.11 & 131.36 & 2.73 & 81.24 & 3.87 & 4.11 & 1.48 & 2.55 \\
J1435-6100 & 9.35 & 113.84 & 1.35 & 83.12 & 3.19 & 3.26 & 1.33 & 2.53 \\
J1446-4701 & 2.19 & 55.83 & 0.28 & 41.51 & 1.61 & 1.81 & 0.94 & 2.55 \\
J1455-3330 & 7.99 & 13.57 & 76.17 & 135.48 & 3.94 & 5.31 & 4.51 & 2.34 \\
J1525-5545 & 11.36 & 127.01 & 0.99 & 155.46 & 3.22 & 3.61 & 2.06 & 2.44 \\
J1543-5149 & 2.06 & 50.98 & 8.06 & 118.19 & 4.02 & 4.72 & 2.69 & 2.53 \\
J1545-4550 & 3.58 & 68.40 & 6.20 & 41.35 & 1.14 & 1.22 & 0.67 & 2.48 \\
J1547-5709 & 4.29 & 95.75 & 3.08 & 75.09 & 4.04 & 4.28 & 1.70 & 2.37 \\
J1600-3053 & 3.60 & 52.32 & 14.35 & 57.91 & 0.48 & 0.54 & 0.24 & 2.24 \\
J1603-7202 & 14.84 & 38.05 & 6.31 & 150.99 & 1.66 & 2.59 & 2.84 & 2.56 \\
J1614-2230 & 3.15 & 34.49 & 8.69 & 53.83 & 1.06 & 1.32 & 0.84 & 2.35 \\
J1629-6902 & 6.00 & 29.50 & - & 93.68 & 1.61 & 2.15 & 1.61 & 2.56 \\
J1643-1224 & 4.62 & 62.40 & 147.02 & 214.91 & 1.16 & 1.21 & 0.39 & 2.45 \\
J1652-4838 & 3.79 & 188.14 & 12.40 & 90.25 & 2.57 & 3.08 & 1.59 & 2.43 \\
J1653-2054 & 4.13 & 56.53 & 1.23 & 129.23 & 4.20 & 4.94 & 2.92 & 2.12 \\
J1658-5324 & 2.44 & 30.83 & - & 27.49 & 2.19 & 2.49 & 1.46 & 2.54 \\
J1705-1903 & 2.48 & 57.51 & - & 26.80 & 0.34 & 0.45 & 0.38 & 2.16 \\
J1708-3506 & 4.51 & 146.81 & 149.13 & 270.90 & 2.99 & 3.67 & 2.35 & 2.45 \\
J1713+0747 & 4.57 & 15.98 & 67.83 & 71.27 & 0.28 & 0.43 & 0.49 & 1.98 \\
J1719-1438 & 5.79 & 36.77 & 0.09 & 128.90 & 4.10 & 5.11 & 3.56 & 2.43 \\
J1721-2457 & 3.50 & 48.22 & - & 259.96 & 4.15 & 4.53 & 1.92 & 2.12 \\
J1730-2304 & 8.12 & 9.63 & - & 110.14 & 1.74 & 2.39 & 2.29 & 2.24 \\
J1732-5049 & 5.31 & 56.82 & 5.26 & 175.35 & 2.74 & 3.93 & 3.58 & 2.62 \\
J1737-0811 & 4.18 & 55.30 & 79.52 & 202.17 & 4.37 & 4.82 & 1.97 & 2.43 \\
J1744-1134 & 4.07 & 3.14 & - & 64.36 & 0.57 & 1.07 & 1.35 & 2.43 \\
J1747-4036 & 1.65 & 152.94 & - & 60.42 & 2.15 & 2.18 & 0.72 & 2.54 \\
J1751-2857 & 3.91 & 42.78 & 110.75 & 79.80 & 3.42 & 3.80 & 1.63 & 2.43 \\
J1756-2251 & 28.46 & 121.24 & 0.32 & 314.28 & 3.47 & 4.11 & 2.00 & 2.50 \\
J1757-5322 & 8.87 & 30.79 & 0.45 & 176.15 & 2.01 & 2.54 & 1.93 & 2.62 \\
J1801-1417 & 3.63 & 57.25 & - & 147.84 & 2.89 & 3.22 & 1.54 & 2.43 \\
J1802-2124 & 12.65 & 149.56 & 0.70 & 102.46 & 1.98 & 2.27 & 1.00 & 2.56 \\
J1811-2405 & 2.66 & 60.64 & 6.27 & 42.69 & 1.04 & 1.03 & 0.47 & 2.49 \\
J1825-0319 & 4.55 & 119.55 & 52.63 & 69.48 & 3.27 & 3.60 & 1.41 & 2.43 \\
J1832-0836 & 2.72 & 28.19 & - & 13.12 & 1.33 & 1.47 & 0.73 & 2.45 \\
J1843-1113 & 1.85 & 59.95 & - & 39.64 & 1.50 & 1.63 & 0.75 & 2.43 \\
J1902-5105 & 1.74 & 36.25 & 2.01 & 50.62 & 1.18 & 1.39 & 0.86 & 2.43 \\
J1903-7051 & 3.60 & 19.65 & 11.05 & 93.02 & 1.94 & 2.89 & 2.68 & 2.54 \\
J1909-3744 & 2.95 & 10.39 & 1.53 & 24.18 & 0.16 & 0.29 & 0.37 & 2.70 \\
J1918-0642 & 7.65 & 26.59 & 10.91 & 105.90 & 1.45 & 1.86 & 1.29 & 2.32 \\
J1933-6211 & 3.54 & 11.51 & 12.82 & 80.84 & 2.55 & 3.84 & 3.95 & 2.56 \\
J1946-5403 & 2.71 & 23.72 & 0.13 & 30.78 & 1.37 & 1.51 & 0.92 & 2.37 \\
J2010-1323 & 5.22 & 22.16 & - & 54.57 & 1.69 & 2.16 & 1.51 & 2.08 \\
J2039-3616 & 3.28 & 23.96 & 5.79 & 76.01 & 2.55 & 3.19 & 2.33 & 2.37 \\
J2124-3358 & 4.93 & 4.60 & - & 141.28 & 2.12 & 2.83 & 2.66 & 2.43 \\
J2129-5721 & 3.73 & 31.85 & 6.63 & 74.06 & 1.99 & 2.99 & 2.85 & 2.49 \\
J2145-0750 & 16.05 & 9.00 & 6.84 & 207.11 & 1.28 & 2.77 & 3.85 & 2.43 \\
J2150-0326 & 3.51 & 20.67 & 4.04 & 58.70 & 2.23 & 2.77 & 1.89 & 2.19 \\
J2222-0137 & 32.82 & 3.27 & 2.45 & 188.84 & 2.56 & 4.31 & 4.73 & 2.19 \\
J2229+2643 & 2.98 & 22.72 & 93.02 & 126.36 & 4.64 & 5.86 & 4.67 & 1.91 \\
J2234+0944 & 3.63 & 17.82 & 0.42 & 88.97 & 2.50 & 3.52 & 2.95 & 1.91 \\
J2241-5236 & 2.19 & 11.41 & - & 26.79 & 0.21 & 0.29 & 0.29 & 2.56 \\
J2317+1439 & 3.45 & 21.90 & 2.46 & 44.77 & 3.17 & 4.06 & 3.15 & 1.91 \\
J2322+2057 & 4.81 & 13.39 & - & 74.36 & 3.53 & 4.33 & 3.04 & 1.91 \\
J2322-2650 & 3.46 & 6.15 & 0.32 & 43.77 & 2.45 & 2.76 & 1.50 & 2.43 \\

\end{longtable}
\twocolumn

\section{Static timing jumps}

\begin{table}

\caption{Known offests that are corrected for in the MPTA.  A positive sign ($+$) indicates a time that needs to be {\em added} to raw MeerKAT/PTUSE data, observed in 1024 channel mode with the L-band system.}
\label{table: jump_table}
\resizebox{\columnwidth}{!}{%
\begin{tabular}{@{}crl@{}}
\toprule
Date Range (MJD)  & Jump Value ($\mu$s) \\ \midrule
$>58526.21089$    & $-24.629$ & Precise Time Manager Time (PTM) not applied in PTUSE observations.           \\
$> 58550.14921$   & $+24.630$ & PTM applied to PTUSE.           \\
$> 58550.14921$   & $-1.196$ &  Half sample offset in PTUSE.\\
$> 58557.14847$   & $-4.785$ & PTM sensor  delay changed.             \\
$>58575.95951$    & $+0.598$ & PTUSE offset changed by half sample.            \\
$58550$ - $58690$ & $-306.243$  & Offset in MeerKAT Correlator Beamformer (CBF)              \\ \bottomrule
\end{tabular}%
}
\end{table}



\bsp	
\label{lastpage}
\end{document}


\begin{figure*} 
	 \centerline{\includegraphics[width=0.9\paperwidth]{Appendix_figs/J0030+0451_grand.dly_4_notebook_comparison.pdf}} 
	 \vspace{-0.5cm} 
	 \centerline{\includegraphics[width=\paperwidth]{Appendix_figs/J0030+0451_noise_reduce_steps.pdf}} 
	 \caption[]{Same as Figure D1, for PSR J0030+0451} 
	 \label{fig: app_J0030+0451} 
	 \vspace{-15pt} 
\end{figure*} 

\begin{figure*} 
	 \centerline{\includegraphics[width=0.9\paperwidth]{Appendix_figs/J0125-2327_grand.dly_4_notebook_comparison.pdf}} 
	 \vspace{-0.5cm} 
	 \centerline{\includegraphics[width=\paperwidth]{Appendix_figs/J0125-2327_noise_reduce_steps.pdf}} 
	 \caption[]{Same as Figure D1, for PSR J0125-2327} 
	 \label{fig: app_J0125-2327} 
	 \vspace{-15pt} 
\end{figure*} 

\begin{figure*} 
	 \centerline{\includegraphics[width=0.9\paperwidth]{Appendix_figs/J0437-4715_grand.dly_4_notebook_comparison.pdf}} 
	 \vspace{-0.5cm} 
	 \centerline{\includegraphics[width=\paperwidth]{Appendix_figs/J0437-4715_noise_reduce_steps.pdf}} 
	 \caption[]{Same as Figure D1, for PSR J0437-4715} 
	 \label{fig: app_J0437-4715} 
	 \vspace{-15pt} 
\end{figure*} 

\begin{figure*} 
	 \centerline{\includegraphics[width=0.9\paperwidth]{Appendix_figs/J0610-2100_grand.dly_4_notebook_comparison.pdf}} 
	 \vspace{-0.5cm} 
	 \centerline{\includegraphics[width=\paperwidth]{Appendix_figs/J0610-2100_noise_reduce_steps.pdf}} 
	 \caption[]{Same as Figure D1, for PSR J0610-2100} 
	 \label{fig: app_J0610-2100} 
	 \vspace{-15pt} 
\end{figure*} 

\begin{figure*} 
	 \centerline{\includegraphics[width=0.9\paperwidth]{Appendix_figs/J0613-0200_grand.dly_4_notebook_comparison.pdf}} 
	 \vspace{-0.5cm} 
	 \centerline{\includegraphics[width=\paperwidth]{Appendix_figs/J0613-0200_noise_reduce_steps.pdf}} 
	 \caption[]{Same as Figure D1, for PSR J0613-0200} 
	 \label{fig: app_J0613-0200} 
	 \vspace{-15pt} 
\end{figure*} 

\begin{figure*} 
	 \centerline{\includegraphics[width=0.9\paperwidth]{Appendix_figs/J0614-3329_grand.dly_4_notebook_comparison.pdf}} 
	 \vspace{-0.5cm} 
	 \centerline{\includegraphics[width=\paperwidth]{Appendix_figs/J0614-3329_noise_reduce_steps.pdf}} 
	 \caption[]{Same as Figure D1, for PSR J0614-3329} 
	 \label{fig: app_J0614-3329} 
	 \vspace{-15pt} 
\end{figure*} 

\begin{figure*} 
	 \centerline{\includegraphics[width=0.9\paperwidth]{Appendix_figs/J0636-3044_grand.dly_4_notebook_comparison.pdf}} 
	 \vspace{-0.5cm} 
	 \centerline{\includegraphics[width=\paperwidth]{Appendix_figs/J0636-3044_noise_reduce_steps.pdf}} 
	 \caption[]{Same as Figure D1, for PSR J0636-3044} 
	 \label{fig: app_J0636-3044} 
	 \vspace{-15pt} 
\end{figure*} 

\begin{figure*} 
	 \centerline{\includegraphics[width=0.9\paperwidth]{Appendix_figs/J0711-6830_grand.dly_4_notebook_comparison.pdf}} 
	 \vspace{-0.5cm} 
	 \centerline{\includegraphics[width=\paperwidth]{Appendix_figs/J0711-6830_noise_reduce_steps.pdf}} 
	 \caption[]{Same as Figure D1, for PSR J0711-6830} 
	 \label{fig: app_J0711-6830} 
	 \vspace{-15pt} 
\end{figure*} 

\begin{figure*} 
	 \centerline{\includegraphics[width=0.9\paperwidth]{Appendix_figs/J0900-3144_grand.dly_4_notebook_comparison.pdf}} 
	 \vspace{-0.5cm} 
	 \centerline{\includegraphics[width=\paperwidth]{Appendix_figs/J0900-3144_noise_reduce_steps.pdf}} 
	 \caption[]{Same as Figure D1, for PSR J0900-3144} 
	 \label{fig: app_J0900-3144} 
	 \vspace{-15pt} 
\end{figure*} 

\begin{figure*} 
	 \centerline{\includegraphics[width=0.9\paperwidth]{Appendix_figs/J0931-1902_grand.dly_4_notebook_comparison.pdf}} 
	 \vspace{-0.5cm} 
	 \centerline{\includegraphics[width=\paperwidth]{Appendix_figs/J0931-1902_noise_reduce_steps.pdf}} 
	 \caption[]{Same as Figure D1, for PSR J0931-1902} 
	 \label{fig: app_J0931-1902} 
	 \vspace{-15pt} 
\end{figure*} 

\begin{figure*} 
	 \centerline{\includegraphics[width=0.9\paperwidth]{Appendix_figs/J0955-6150_grand.dly_4_notebook_comparison.pdf}} 
	 \vspace{-0.5cm} 
	 \centerline{\includegraphics[width=\paperwidth]{Appendix_figs/J0955-6150_noise_reduce_steps.pdf}} 
	 \caption[]{Same as Figure D1, for PSR J0955-6150} 
	 \label{fig: app_J0955-6150} 
	 \vspace{-15pt} 
\end{figure*} 

\begin{figure*} 
	 \centerline{\includegraphics[width=0.9\paperwidth]{Appendix_figs/J1012-4235_grand.dly_4_notebook_comparison.pdf}} 
	 \vspace{-0.5cm} 
	 \centerline{\includegraphics[width=\paperwidth]{Appendix_figs/J1012-4235_noise_reduce_steps.pdf}} 
	 \caption[]{Same as Figure D1, for PSR J1012-4235} 
	 \label{fig: app_J1012-4235} 
	 \vspace{-15pt} 
\end{figure*} 

\begin{figure*} 
	 \centerline{\includegraphics[width=0.9\paperwidth]{Appendix_figs/J1017-7156_grand.dly_4_notebook_comparison.pdf}} 
	 \vspace{-0.5cm} 
	 \centerline{\includegraphics[width=\paperwidth]{Appendix_figs/J1017-7156_noise_reduce_steps.pdf}} 
	 \caption[]{Same as Figure D1, for PSR J1017-7156} 
	 \label{fig: app_J1017-7156} 
	 \vspace{-15pt} 
\end{figure*} 

\begin{figure*} 
	 \centerline{\includegraphics[width=0.9\paperwidth]{Appendix_figs/J1022+1001_grand.dly_4_notebook_comparison.pdf}} 
	 \vspace{-0.5cm} 
	 \centerline{\includegraphics[width=\paperwidth]{Appendix_figs/J1022+1001_noise_reduce_steps.pdf}} 
	 \caption[]{Same as Figure D1, for PSR J1022+1001} 
	 \label{fig: app_J1022+1001} 
	 \vspace{-15pt} 
\end{figure*} 

\begin{figure*} 
	 \centerline{\includegraphics[width=0.9\paperwidth]{Appendix_figs/J1024-0719_grand.dly_4_notebook_comparison.pdf}} 
	 \vspace{-0.5cm} 
	 \centerline{\includegraphics[width=\paperwidth]{Appendix_figs/J1024-0719_noise_reduce_steps.pdf}} 
	 \caption[]{Same as Figure D1, for PSR J1024-0719} 
	 \label{fig: app_J1024-0719} 
	 \vspace{-15pt} 
\end{figure*} 

\begin{figure*} 
	 \centerline{\includegraphics[width=0.9\paperwidth]{Appendix_figs/J1036-8317_grand.dly_4_notebook_comparison.pdf}} 
	 \vspace{-0.5cm} 
	 \centerline{\includegraphics[width=\paperwidth]{Appendix_figs/J1036-8317_noise_reduce_steps.pdf}} 
	 \caption[]{Same as Figure D1, for PSR J1036-8317} 
	 \label{fig: app_J1036-8317} 
	 \vspace{-15pt} 
\end{figure*} 

\begin{figure*} 
	 \centerline{\includegraphics[width=0.9\paperwidth]{Appendix_figs/J1045-4509_grand.dly_4_notebook_comparison.pdf}} 
	 \vspace{-0.5cm} 
	 \centerline{\includegraphics[width=\paperwidth]{Appendix_figs/J1045-4509_noise_reduce_steps.pdf}} 
	 \caption[]{Same as Figure D1, for PSR J1045-4509} 
	 \label{fig: app_J1045-4509} 
	 \vspace{-15pt} 
\end{figure*} 

\begin{figure*} 
	 \centerline{\includegraphics[width=0.9\paperwidth]{Appendix_figs/J1101-6424_grand.dly_4_notebook_comparison.pdf}} 
	 \vspace{-0.5cm} 
	 \centerline{\includegraphics[width=\paperwidth]{Appendix_figs/J1101-6424_noise_reduce_steps.pdf}} 
	 \caption[]{Same as Figure D1, for PSR J1101-6424} 
	 \label{fig: app_J1101-6424} 
	 \vspace{-15pt} 
\end{figure*} 

\begin{figure*} 
	 \centerline{\includegraphics[width=0.9\paperwidth]{Appendix_figs/J1103-5403_grand.dly_4_notebook_comparison.pdf}} 
	 \vspace{-0.5cm} 
	 \centerline{\includegraphics[width=\paperwidth]{Appendix_figs/J1103-5403_noise_reduce_steps.pdf}} 
	 \caption[]{Same as Figure D1, for PSR J1103-5403} 
	 \label{fig: app_J1103-5403} 
	 \vspace{-15pt} 
\end{figure*} 

\begin{figure*} 
	 \centerline{\includegraphics[width=0.9\paperwidth]{Appendix_figs/J1125-5825_grand.dly_4_notebook_comparison.pdf}} 
	 \vspace{-0.5cm} 
	 \centerline{\includegraphics[width=\paperwidth]{Appendix_figs/J1125-5825_noise_reduce_steps.pdf}} 
	 \caption[]{Same as Figure D1, for PSR J1125-5825} 
	 \label{fig: app_J1125-5825} 
	 \vspace{-15pt} 
\end{figure*} 

\begin{figure*} 
	 \centerline{\includegraphics[width=0.9\paperwidth]{Appendix_figs/J1125-6014_grand.dly_4_notebook_comparison.pdf}} 
	 \vspace{-0.5cm} 
	 \centerline{\includegraphics[width=\paperwidth]{Appendix_figs/J1125-6014_noise_reduce_steps.pdf}} 
	 \caption[]{Same as Figure D1, for PSR J1125-6014} 
	 \label{fig: app_J1125-6014} 
	 \vspace{-15pt} 
\end{figure*} 

\begin{figure*} 
	 \centerline{\includegraphics[width=0.9\paperwidth]{Appendix_figs/J1216-6410_grand.dly_4_notebook_comparison.pdf}} 
	 \vspace{-0.5cm} 
	 \centerline{\includegraphics[width=\paperwidth]{Appendix_figs/J1216-6410_noise_reduce_steps.pdf}} 
	 \caption[]{Same as Figure D1, for PSR J1216-6410} 
	 \label{fig: app_J1216-6410} 
	 \vspace{-15pt} 
\end{figure*} 

\begin{figure*} 
	 \centerline{\includegraphics[width=0.9\paperwidth]{Appendix_figs/J1327-0755_grand.dly_4_notebook_comparison.pdf}} 
	 \vspace{-0.5cm} 
	 \centerline{\includegraphics[width=\paperwidth]{Appendix_figs/J1327-0755_noise_reduce_steps.pdf}} 
	 \caption[]{Same as Figure D1, for PSR J1327-0755} 
	 \label{fig: app_J1327-0755} 
	 \vspace{-15pt} 
\end{figure*} 

\begin{figure*} 
	 \centerline{\includegraphics[width=0.9\paperwidth]{Appendix_figs/J1421-4409_grand.dly_4_notebook_comparison.pdf}} 
	 \vspace{-0.5cm} 
	 \centerline{\includegraphics[width=\paperwidth]{Appendix_figs/J1421-4409_noise_reduce_steps.pdf}} 
	 \caption[]{Same as Figure D1, for PSR J1421-4409} 
	 \label{fig: app_J1421-4409} 
	 \vspace{-15pt} 
\end{figure*} 

\begin{figure*} 
	 \centerline{\includegraphics[width=0.9\paperwidth]{Appendix_figs/J1431-5740_grand.dly_4_notebook_comparison.pdf}} 
	 \vspace{-0.5cm} 
	 \centerline{\includegraphics[width=\paperwidth]{Appendix_figs/J1431-5740_noise_reduce_steps.pdf}} 
	 \caption[]{Same as Figure D1, for PSR J1431-5740} 
	 \label{fig: app_J1431-5740} 
	 \vspace{-15pt} 
\end{figure*} 

\begin{figure*} 
	 \centerline{\includegraphics[width=0.9\paperwidth]{Appendix_figs/J1435-6100_grand.dly_4_notebook_comparison.pdf}} 
	 \vspace{-0.5cm} 
	 \centerline{\includegraphics[width=\paperwidth]{Appendix_figs/J1435-6100_noise_reduce_steps.pdf}} 
	 \caption[]{Same as Figure D1, for PSR J1435-6100} 
	 \label{fig: app_J1435-6100} 
	 \vspace{-15pt} 
\end{figure*} 

\begin{figure*} 
	 \centerline{\includegraphics[width=0.9\paperwidth]{Appendix_figs/J1446-4701_grand.dly_4_notebook_comparison.pdf}} 
	 \vspace{-0.5cm} 
	 \centerline{\includegraphics[width=\paperwidth]{Appendix_figs/J1446-4701_noise_reduce_steps.pdf}} 
	 \caption[]{Same as Figure D1, for PSR J1446-4701} 
	 \label{fig: app_J1446-4701} 
	 \vspace{-15pt} 
\end{figure*} 

\begin{figure*} 
	 \centerline{\includegraphics[width=0.9\paperwidth]{Appendix_figs/J1455-3330_grand.dly_4_notebook_comparison.pdf}} 
	 \vspace{-0.5cm} 
	 \centerline{\includegraphics[width=\paperwidth]{Appendix_figs/J1455-3330_noise_reduce_steps.pdf}} 
	 \caption[]{Same as Figure D1, for PSR J1455-3330} 
	 \label{fig: app_J1455-3330} 
	 \vspace{-15pt} 
\end{figure*} 

\begin{figure*} 
	 \centerline{\includegraphics[width=0.9\paperwidth]{Appendix_figs/J1525-5545_grand.dly_4_notebook_comparison.pdf}} 
	 \vspace{-0.5cm} 
	 \centerline{\includegraphics[width=\paperwidth]{Appendix_figs/J1525-5545_noise_reduce_steps.pdf}} 
	 \caption[]{Same as Figure D1, for PSR J1525-5545} 
	 \label{fig: app_J1525-5545} 
	 \vspace{-15pt} 
\end{figure*} 

\begin{figure*} 
	 \centerline{\includegraphics[width=0.9\paperwidth]{Appendix_figs/J1543-5149_grand.dly_4_notebook_comparison.pdf}} 
	 \vspace{-0.5cm} 
	 \centerline{\includegraphics[width=\paperwidth]{Appendix_figs/J1543-5149_noise_reduce_steps.pdf}} 
	 \caption[]{Same as Figure D1, for PSR J1543-5149} 
	 \label{fig: app_J1543-5149} 
	 \vspace{-15pt} 
\end{figure*} 

\begin{figure*} 
	 \centerline{\includegraphics[width=0.9\paperwidth]{Appendix_figs/J1545-4550_grand.dly_4_notebook_comparison.pdf}} 
	 \vspace{-0.5cm} 
	 \centerline{\includegraphics[width=\paperwidth]{Appendix_figs/J1545-4550_noise_reduce_steps.pdf}} 
	 \caption[]{Same as Figure D1, for PSR J1545-4550} 
	 \label{fig: app_J1545-4550} 
	 \vspace{-15pt} 
\end{figure*} 

\begin{figure*} 
	 \centerline{\includegraphics[width=0.9\paperwidth]{Appendix_figs/J1547-5709_grand.dly_4_notebook_comparison.pdf}} 
	 \vspace{-0.5cm} 
	 \centerline{\includegraphics[width=\paperwidth]{Appendix_figs/J1547-5709_noise_reduce_steps.pdf}} 
	 \caption[]{Same as Figure D1, for PSR J1547-5709} 
	 \label{fig: app_J1547-5709} 
	 \vspace{-15pt} 
\end{figure*} 

\begin{figure*} 
	 \centerline{\includegraphics[width=0.9\paperwidth]{Appendix_figs/J1600-3053_grand.dly_4_notebook_comparison.pdf}} 
	 \vspace{-0.5cm} 
	 \centerline{\includegraphics[width=\paperwidth]{Appendix_figs/J1600-3053_noise_reduce_steps.pdf}} 
	 \caption[]{Same as Figure D1, for PSR J1600-3053} 
	 \label{fig: app_J1600-3053} 
	 \vspace{-15pt} 
\end{figure*} 

\begin{figure*} 
	 \centerline{\includegraphics[width=0.9\paperwidth]{Appendix_figs/J1603-7202_grand.dly_4_notebook_comparison.pdf}} 
	 \vspace{-0.5cm} 
	 \centerline{\includegraphics[width=\paperwidth]{Appendix_figs/J1603-7202_noise_reduce_steps.pdf}} 
	 \caption[]{Same as Figure D1, for PSR J1603-7202} 
	 \label{fig: app_J1603-7202} 
	 \vspace{-15pt} 
\end{figure*} 

\begin{figure*} 
	 \centerline{\includegraphics[width=0.9\paperwidth]{Appendix_figs/J1614-2230_grand.dly_4_notebook_comparison.pdf}} 
	 \vspace{-0.5cm} 
	 \centerline{\includegraphics[width=\paperwidth]{Appendix_figs/J1614-2230_noise_reduce_steps.pdf}} 
	 \caption[]{Same as Figure D1, for PSR J1614-2230} 
	 \label{fig: app_J1614-2230} 
	 \vspace{-15pt} 
\end{figure*} 

\begin{figure*} 
	 \centerline{\includegraphics[width=0.9\paperwidth]{Appendix_figs/J1629-6902_grand.dly_4_notebook_comparison.pdf}} 
	 \vspace{-0.5cm} 
	 \centerline{\includegraphics[width=\paperwidth]{Appendix_figs/J1629-6902_noise_reduce_steps.pdf}} 
	 \caption[]{Same as Figure D1, for PSR J1629-6902} 
	 \label{fig: app_J1629-6902} 
	 \vspace{-15pt} 
\end{figure*} 

\begin{figure*} 
	 \centerline{\includegraphics[width=0.9\paperwidth]{Appendix_figs/J1643-1224_grand.dly_4_notebook_comparison.pdf}} 
	 \vspace{-0.5cm} 
	 \centerline{\includegraphics[width=\paperwidth]{Appendix_figs/J1643-1224_noise_reduce_steps.pdf}} 
	 \caption[]{Same as Figure D1, for PSR J1643-1224} 
	 \label{fig: app_J1643-1224} 
	 \vspace{-15pt} 
\end{figure*} 

\begin{figure*} 
	 \centerline{\includegraphics[width=0.9\paperwidth]{Appendix_figs/J1652-4838_grand.dly_4_notebook_comparison.pdf}} 
	 \vspace{-0.5cm} 
	 \centerline{\includegraphics[width=\paperwidth]{Appendix_figs/J1652-4838_noise_reduce_steps.pdf}} 
	 \caption[]{Same as Figure D1, for PSR J1652-4838} 
	 \label{fig: app_J1652-4838} 
	 \vspace{-15pt} 
\end{figure*} 

\begin{figure*} 
	 \centerline{\includegraphics[width=0.9\paperwidth]{Appendix_figs/J1653-2054_grand.dly_4_notebook_comparison.pdf}} 
	 \vspace{-0.5cm} 
	 \centerline{\includegraphics[width=\paperwidth]{Appendix_figs/J1653-2054_noise_reduce_steps.pdf}} 
	 \caption[]{Same as Figure D1, for PSR J1653-2054} 
	 \label{fig: app_J1653-2054} 
	 \vspace{-15pt} 
\end{figure*} 

\begin{figure*} 
	 \centerline{\includegraphics[width=0.9\paperwidth]{Appendix_figs/J1658-5324_grand.dly_4_notebook_comparison.pdf}} 
	 \vspace{-0.5cm} 
	 \centerline{\includegraphics[width=\paperwidth]{Appendix_figs/J1658-5324_noise_reduce_steps.pdf}} 
	 \caption[]{Same as Figure D1, for PSR J1658-5324} 
	 \label{fig: app_J1658-5324} 
	 \vspace{-15pt} 
\end{figure*} 

\begin{figure*} 
	 \centerline{\includegraphics[width=0.9\paperwidth]{Appendix_figs/J1705-1903_grand.dly_4_notebook_comparison.pdf}} 
	 \vspace{-0.5cm} 
	 \centerline{\includegraphics[width=\paperwidth]{Appendix_figs/J1705-1903_noise_reduce_steps.pdf}} 
	 \caption[]{Same as Figure D1, for PSR J1705-1903} 
	 \label{fig: app_J1705-1903} 
	 \vspace{-15pt} 
\end{figure*} 

\begin{figure*} 
	 \centerline{\includegraphics[width=0.9\paperwidth]{Appendix_figs/J1708-3506_grand.dly_4_notebook_comparison.pdf}} 
	 \vspace{-0.5cm} 
	 \centerline{\includegraphics[width=\paperwidth]{Appendix_figs/J1708-3506_noise_reduce_steps.pdf}} 
	 \caption[]{Same as Figure D1, for PSR J1708-3506} 
	 \label{fig: app_J1708-3506} 
	 \vspace{-15pt} 
\end{figure*} 

\begin{figure*} 
	 \centerline{\includegraphics[width=0.9\paperwidth]{Appendix_figs/J1713+0747_grand.dly_4_notebook_comparison.pdf}} 
	 \vspace{-0.5cm} 
	 \centerline{\includegraphics[width=\paperwidth]{Appendix_figs/J1713+0747_noise_reduce_steps.pdf}} 
	 \caption[]{Same as Figure D1, for PSR J1713+0747} 
	 \label{fig: app_J1713+0747} 
	 \vspace{-15pt} 
\end{figure*} 

\begin{figure*} 
	 \centerline{\includegraphics[width=0.9\paperwidth]{Appendix_figs/J1719-1438_grand.dly_4_notebook_comparison.pdf}} 
	 \vspace{-0.5cm} 
	 \centerline{\includegraphics[width=\paperwidth]{Appendix_figs/J1719-1438_noise_reduce_steps.pdf}} 
	 \caption[]{Same as Figure D1, for PSR J1719-1438} 
	 \label{fig: app_J1719-1438} 
	 \vspace{-15pt} 
\end{figure*} 

\begin{figure*} 
	 \centerline{\includegraphics[width=0.9\paperwidth]{Appendix_figs/J1721-2457_grand.dly_4_notebook_comparison.pdf}} 
	 \vspace{-0.5cm} 
	 \centerline{\includegraphics[width=\paperwidth]{Appendix_figs/J1721-2457_noise_reduce_steps.pdf}} 
	 \caption[]{Same as Figure D1, for PSR J1721-2457} 
	 \label{fig: app_J1721-2457} 
	 \vspace{-15pt} 
\end{figure*} 

\begin{figure*} 
	 \centerline{\includegraphics[width=0.9\paperwidth]{Appendix_figs/J1730-2304_grand.dly_4_notebook_comparison.pdf}} 
	 \vspace{-0.5cm} 
	 \centerline{\includegraphics[width=\paperwidth]{Appendix_figs/J1730-2304_noise_reduce_steps.pdf}} 
	 \caption[]{Same as Figure D1, for PSR J1730-2304} 
	 \label{fig: app_J1730-2304} 
	 \vspace{-15pt} 
\end{figure*} 

\begin{figure*} 
	 \centerline{\includegraphics[width=0.9\paperwidth]{Appendix_figs/J1732-5049_grand.dly_4_notebook_comparison.pdf}} 
	 \vspace{-0.5cm} 
	 \centerline{\includegraphics[width=\paperwidth]{Appendix_figs/J1732-5049_noise_reduce_steps.pdf}} 
	 \caption[]{Same as Figure D1, for PSR J1732-5049} 
	 \label{fig: app_J1732-5049} 
	 \vspace{-15pt} 
\end{figure*} 

\begin{figure*} 
	 \centerline{\includegraphics[width=0.9\paperwidth]{Appendix_figs/J1737-0811_grand.dly_4_notebook_comparison.pdf}} 
	 \vspace{-0.5cm} 
	 \centerline{\includegraphics[width=\paperwidth]{Appendix_figs/J1737-0811_noise_reduce_steps.pdf}} 
	 \caption[]{Same as Figure D1, for PSR J1737-0811} 
	 \label{fig: app_J1737-0811} 
	 \vspace{-15pt} 
\end{figure*} 
\afterpage{\clearpage}

\begin{figure*} 
	 \centerline{\includegraphics[width=0.9\paperwidth]{Appendix_figs/J1744-1134_grand.dly_4_notebook_comparison.pdf}} 
	 \vspace{-0.5cm} 
	 \centerline{\includegraphics[width=\paperwidth]{Appendix_figs/J1744-1134_noise_reduce_steps.pdf}} 
	 \caption[]{Same as Figure D1, for PSR J1744-1134} 
	 \label{fig: app_J1744-1134} 
	 \vspace{-15pt} 
\end{figure*} 

\begin{figure*} 
	 \centerline{\includegraphics[width=0.9\paperwidth]{Appendix_figs/J1747-4036_grand.dly_4_notebook_comparison.pdf}} 
	 \vspace{-0.5cm} 
	 \centerline{\includegraphics[width=\paperwidth]{Appendix_figs/J1747-4036_noise_reduce_steps.pdf}} 
	 \caption[]{Same as Figure D1, for PSR J1747-4036} 
	 \label{fig: app_J1747-4036} 
	 \vspace{-15pt} 
\end{figure*} 

\begin{figure*} 
	 \centerline{\includegraphics[width=0.9\paperwidth]{Appendix_figs/J1751-2857_grand.dly_4_notebook_comparison.pdf}} 
	 \vspace{-0.5cm} 
	 \centerline{\includegraphics[width=\paperwidth]{Appendix_figs/J1751-2857_noise_reduce_steps.pdf}} 
	 \caption[]{Same as Figure D1, for PSR J1751-2857} 
	 \label{fig: app_J1751-2857} 
	 \vspace{-15pt} 
\end{figure*}

\begin{figure*} 
	 \centerline{\includegraphics[width=0.9\paperwidth]{Appendix_figs/J1756-2251_grand.dly_4_notebook_comparison.pdf}} 
	 \vspace{-0.5cm} 
	 \centerline{\includegraphics[width=\paperwidth]{Appendix_figs/J1756-2251_noise_reduce_steps.pdf}} 
	 \caption[]{Same as Figure D1, for PSR J1756-2251} 
	 \label{fig: app_J1756-2251} 
	 \vspace{-15pt} 
\end{figure*} 

\begin{figure*} 
	 \centerline{\includegraphics[width=0.9\paperwidth]{Appendix_figs/J1757-5322_grand.dly_4_notebook_comparison.pdf}} 
	 \vspace{-0.5cm} 
	 \centerline{\includegraphics[width=\paperwidth]{Appendix_figs/J1757-5322_noise_reduce_steps.pdf}} 
	 \caption[]{Same as Figure D1, for PSR J1757-5322} 
	 \label{fig: app_J1757-5322} 
	 \vspace{-15pt} 
\end{figure*} 

\begin{figure*} 
	 \centerline{\includegraphics[width=0.9\paperwidth]{Appendix_figs/J1801-1417_grand.dly_4_notebook_comparison.pdf}} 
	 \vspace{-0.5cm} 
	 \centerline{\includegraphics[width=\paperwidth]{Appendix_figs/J1801-1417_noise_reduce_steps.pdf}} 
	 \caption[]{Same as Figure D1, for PSR J1801-1417} 
	 \label{fig: app_J1801-1417} 
	 \vspace{-15pt} 
\end{figure*} 

\begin{figure*} 
	 \centerline{\includegraphics[width=0.9\paperwidth]{Appendix_figs/J1802-2124_grand.dly_4_notebook_comparison.pdf}} 
	 \vspace{-0.5cm} 
	 \centerline{\includegraphics[width=\paperwidth]{Appendix_figs/J1802-2124_noise_reduce_steps.pdf}} 
	 \caption[]{Same as Figure D1, for PSR J1802-2124} 
	 \label{fig: app_J1802-2124} 
	 \vspace{-15pt} 
\end{figure*} 

\begin{figure*} 
	 \centerline{\includegraphics[width=0.9\paperwidth]{Appendix_figs/J1811-2405_grand.dly_4_notebook_comparison.pdf}} 
	 \vspace{-0.5cm} 
	 \centerline{\includegraphics[width=\paperwidth]{Appendix_figs/J1811-2405_noise_reduce_steps.pdf}} 
	 \caption[]{Same as Figure D1, for PSR J1811-2405} 
	 \label{fig: app_J1811-2405} 
	 \vspace{-15pt} 
\end{figure*} 

\begin{figure*} 
	 \centerline{\includegraphics[width=0.9\paperwidth]{Appendix_figs/J1825-0319_grand.dly_4_notebook_comparison.pdf}} 
	 \vspace{-0.5cm} 
	 \centerline{\includegraphics[width=\paperwidth]{Appendix_figs/J1825-0319_noise_reduce_steps.pdf}} 
	 \caption[]{Same as Figure D1, for PSR J1825-0319} 
	 \label{fig: app_J1825-0319} 
	 \vspace{-15pt} 
\end{figure*} 

\begin{figure*} 
	 \centerline{\includegraphics[width=0.9\paperwidth]{Appendix_figs/J1832-0836_grand.dly_4_notebook_comparison.pdf}} 
	 \vspace{-0.5cm} 
	 \centerline{\includegraphics[width=\paperwidth]{Appendix_figs/J1832-0836_noise_reduce_steps.pdf}} 
	 \caption[]{Same as Figure D1, for PSR J1832-0836} 
	 \label{fig: app_J1832-0836} 
	 \vspace{-15pt} 
\end{figure*} 

\begin{figure*} 
	 \centerline{\includegraphics[width=0.9\paperwidth]{Appendix_figs/J1843-1113_grand.dly_4_notebook_comparison.pdf}} 
	 \vspace{-0.5cm} 
	 \centerline{\includegraphics[width=\paperwidth]{Appendix_figs/J1843-1113_noise_reduce_steps.pdf}} 
	 \caption[]{Same as Figure D1, for PSR J1843-1113} 
	 \label{fig: app_J1843-1113} 
	 \vspace{-15pt} 
\end{figure*} 

\begin{figure*} 
	 \centerline{\includegraphics[width=0.9\paperwidth]{Appendix_figs/J1902-5105_grand.dly_4_notebook_comparison.pdf}} 
	 \vspace{-0.5cm} 
	 \centerline{\includegraphics[width=\paperwidth]{Appendix_figs/J1902-5105_noise_reduce_steps.pdf}} 
	 \caption[]{Same as Figure D1, for PSR J1902-5105} 
	 \label{fig: app_J1902-5105} 
	 \vspace{-15pt} 
\end{figure*} 

\begin{figure*} 
	 \centerline{\includegraphics[width=0.9\paperwidth]{Appendix_figs/J1903-7051_grand.dly_4_notebook_comparison.pdf}} 
	 \vspace{-0.5cm} 
	 \centerline{\includegraphics[width=\paperwidth]{Appendix_figs/J1903-7051_noise_reduce_steps.pdf}} 
	 \caption[]{Same as Figure D1, for PSR J1903-7051} 
	 \label{fig: app_J1903-7051} 
	 \vspace{-15pt} 
\end{figure*} 

\begin{figure*} 
	 \centerline{\includegraphics[width=0.9\paperwidth]{Appendix_figs/J1909-3744_grand.dly_4_notebook_comparison.pdf}} 
	 \vspace{-0.5cm} 
	 \centerline{\includegraphics[width=\paperwidth]{Appendix_figs/J1909-3744_noise_reduce_steps.pdf}} 
	 \caption[]{Same as Figure D1, for PSR J1909-3744} 
	 \label{fig: app_J1909-3744} 
	 \vspace{-15pt} 
\end{figure*} 

\begin{figure*} 
	 \centerline{\includegraphics[width=0.9\paperwidth]{Appendix_figs/J1918-0642_grand.dly_4_notebook_comparison.pdf}} 
	 \vspace{-0.5cm} 
	 \centerline{\includegraphics[width=\paperwidth]{Appendix_figs/J1918-0642_noise_reduce_steps.pdf}} 
	 \caption[]{Same as Figure D1, for PSR J1918-0642} 
	 \label{fig: app_J1918-0642} 
	 \vspace{-15pt} 
\end{figure*} 

\begin{figure*} 
	 \centerline{\includegraphics[width=0.9\paperwidth]{Appendix_figs/J1933-6211_grand.dly_4_notebook_comparison.pdf}} 
	 \vspace{-0.5cm} 
	 \centerline{\includegraphics[width=\paperwidth]{Appendix_figs/J1933-6211_noise_reduce_steps.pdf}} 
	 \caption[]{Same as Figure D1, for PSR J1933-6211} 
	 \label{fig: app_J1933-6211} 
	 \vspace{-15pt} 
\end{figure*} 

\begin{figure*} 
	 \centerline{\includegraphics[width=0.9\paperwidth]{Appendix_figs/J1946-5403_grand.dly_4_notebook_comparison.pdf}} 
	 \vspace{-0.5cm} 
	 \centerline{\includegraphics[width=\paperwidth]{Appendix_figs/J1946-5403_noise_reduce_steps.pdf}} 
	 \caption[]{Same as Figure D1, for PSR J1946-5403} 
	 \label{fig: app_J1946-5403} 
	 \vspace{-15pt} 
\end{figure*} 

\begin{figure*} 
	 \centerline{\includegraphics[width=0.9\paperwidth]{Appendix_figs/J2010-1323_grand.dly_4_notebook_comparison.pdf}} 
	 \vspace{-0.5cm} 
	 \centerline{\includegraphics[width=\paperwidth]{Appendix_figs/J2010-1323_noise_reduce_steps.pdf}} 
	 \caption[]{Same as Figure D1, for PSR J2010-1323} 
	 \label{fig: app_J2010-1323} 
	 \vspace{-15pt} 
\end{figure*} 

\begin{figure*} 
	 \centerline{\includegraphics[width=0.9\paperwidth]{Appendix_figs/J2039-3616_grand.dly_4_notebook_comparison.pdf}} 
	 \vspace{-0.5cm} 
	 \centerline{\includegraphics[width=\paperwidth]{Appendix_figs/J2039-3616_noise_reduce_steps.pdf}} 
	 \caption[]{Same as Figure D1, for PSR J2039-3616} 
	 \label{fig: app_J2039-3616} 
	 \vspace{-15pt} 
\end{figure*} 

\begin{figure*} 
	 \centerline{\includegraphics[width=0.9\paperwidth]{Appendix_figs/J2124-3358_grand.dly_4_notebook_comparison.pdf}} 
	 \vspace{-0.5cm} 
	 \centerline{\includegraphics[width=\paperwidth]{Appendix_figs/J2124-3358_noise_reduce_steps.pdf}} 
	 \caption[]{Same as Figure D1, for PSR J2124-3358} 
	 \label{fig: app_J2124-3358} 
	 \vspace{-15pt} 
\end{figure*} 

\begin{figure*} 
	 \centerline{\includegraphics[width=0.9\paperwidth]{Appendix_figs/J2129-5721_grand.dly_4_notebook_comparison.pdf}} 
	 \vspace{-0.5cm} 
	 \centerline{\includegraphics[width=\paperwidth]{Appendix_figs/J2129-5721_noise_reduce_steps.pdf}} 
	 \caption[]{Same as Figure D1, for PSR J2129-5721} 
	 \label{fig: app_J2129-5721} 
	 \vspace{-15pt} 
\end{figure*} 

\begin{figure*} 
	 \centerline{\includegraphics[width=0.9\paperwidth]{Appendix_figs/J2145-0750_grand.dly_4_notebook_comparison.pdf}} 
	 \vspace{-0.5cm} 
	 \centerline{\includegraphics[width=\paperwidth]{Appendix_figs/J2145-0750_noise_reduce_steps.pdf}} 
	 \caption[]{Same as Figure D1, for PSR J2145-0750} 
	 \label{fig: app_J2145-0750} 
	 \vspace{-15pt} 
\end{figure*} 

\begin{figure*} 
	 \centerline{\includegraphics[width=0.9\paperwidth]{Appendix_figs/J2150-0326_grand.dly_4_notebook_comparison.pdf}} 
	 \vspace{-0.5cm} 
	 \centerline{\includegraphics[width=\paperwidth]{Appendix_figs/J2150-0326_noise_reduce_steps.pdf}} 
	 \caption[]{Same as Figure D1, for PSR J2150-0326} 
	 \label{fig: app_J2150-0326} 
	 \vspace{-15pt} 
\end{figure*} 

\begin{figure*} 
	 \centerline{\includegraphics[width=0.9\paperwidth]{Appendix_figs/J2222-0137_grand.dly_4_notebook_comparison.pdf}} 
	 \vspace{-0.5cm} 
	 \centerline{\includegraphics[width=\paperwidth]{Appendix_figs/J2222-0137_noise_reduce_steps.pdf}} 
	 \caption[]{Same as Figure D1, for PSR J2222-0137} 
	 \label{fig: app_J2222-0137} 
	 \vspace{-15pt} 
\end{figure*} 

\begin{figure*} 
	 \centerline{\includegraphics[width=0.9\paperwidth]{Appendix_figs/J2229+2643_grand.dly_4_notebook_comparison.pdf}} 
	 \vspace{-0.5cm} 
	 \centerline{\includegraphics[width=\paperwidth]{Appendix_figs/J2229+2643_noise_reduce_steps.pdf}} 
	 \caption[]{Same as Figure D1, for PSR J2229+2643} 
	 \label{fig: app_J2229+2643} 
	 \vspace{-15pt} 
\end{figure*} 

\begin{figure*} 
	 \centerline{\includegraphics[width=0.9\paperwidth]{Appendix_figs/J2234+0944_grand.dly_4_notebook_comparison.pdf}} 
	 \vspace{-0.5cm} 
	 \centerline{\includegraphics[width=\paperwidth]{Appendix_figs/J2234+0944_noise_reduce_steps.pdf}} 
	 \caption[]{Same as Figure D1, for PSR J2234+0944} 
	 \label{fig: app_J2234+0944} 
	 \vspace{-15pt} 
\end{figure*} 

\begin{figure*} 
	 \centerline{\includegraphics[width=0.9\paperwidth]{Appendix_figs/J2241-5236_grand.dly_4_notebook_comparison.pdf}} 
	 \vspace{-0.5cm} 
	 \centerline{\includegraphics[width=\paperwidth]{Appendix_figs/J2241-5236_noise_reduce_steps.pdf}} 
	 \caption[]{Same as Figure D1, for PSR J2241-5236} 
	 \label{fig: app_J2241-5236} 
	 \vspace{-15pt} 
\end{figure*} 

\begin{figure*} 
	 \centerline{\includegraphics[width=0.9\paperwidth]{Appendix_figs/J2317+1439_grand.dly_4_notebook_comparison.pdf}} 
	 \vspace{-0.5cm} 
	 \centerline{\includegraphics[width=\paperwidth]{Appendix_figs/J2317+1439_noise_reduce_steps.pdf}} 
	 \caption[]{Same as Figure D1, for PSR J2317+1439} 
	 \label{fig: app_J2317+1439} 
	 \vspace{-15pt} 
\end{figure*} 

\begin{figure*} 
	 \centerline{\includegraphics[width=0.9\paperwidth]{Appendix_figs/J2322+2057_grand.dly_4_notebook_comparison.pdf}} 
	 \vspace{-0.5cm} 
	 \centerline{\includegraphics[width=\paperwidth]{Appendix_figs/J2322+2057_noise_reduce_steps.pdf}} 
	 \caption[]{Same as Figure D1, for PSR J2322+2057} 
	 \label{fig: app_J2322+2057} 
	 \vspace{-15pt} 
\end{figure*} 

\begin{figure*} 
	 \centerline{\includegraphics[width=0.9\paperwidth]{Appendix_figs/J2322-2650_grand.dly_4_notebook_comparison.pdf}} 
	 \vspace{-0.5cm} 
	 \centerline{\includegraphics[width=\paperwidth]{Appendix_figs/J2322-2650_noise_reduce_steps.pdf}} 
	 \caption[]{Same as Figure D1, for PSR J2322-2650} 
	 \label{fig: app_J2322-2650} 
	 \vspace{-15pt} 
\end{figure*}